\documentclass[usenatbib,fleqn]{mnras}
\usepackage{graphicx}
\usepackage{amssymb}
\usepackage{amsmath}

\newcommand{\lcur}{\mathcal{L}}

\newcommand{\Mcur}{\mathcal{M}}
\newcommand{\Fcur}{\mathcal{F}}

\newcommand{\Zcur}{\mathcal{Z}}

\usepackage[usenames,dvipsnames]{xcolor}

\newcommand{\ba}{\begin{eqnarray}}
\newcommand{\ea}{\end{eqnarray}}

\newcommand{\blue}[1]{\textcolor{blue}{#1}}
\newcommand\boldblue[1]{\textcolor{blue}{\textbf{#1}}}

\title[Phenomenological $P(k)$ models for Roman]
{Phenomenological power spectrum models for H$\alpha$ emission line galaxies from the Nancy Grace Roman Space Telescope}

\author[McCarthy, Zhai, \& Wang]{
Kevin S. McCarthy$^{1}$\thanks{kevin.s.mccarthy@jpl.nasa.gov; NASA Postdoctoral Fellow},
Zhongxu Zhai$^{2,3,4,5}$,
and Yun Wang$^{6}$\\
$^{1}$ Jet Propulsion Laboratory, California Institute of Technology, Pasadena, CA 91109, USA\\
$^{2}$ Department of Astronomy, School of Physics and Astronomy, Shanghai Jiao Tong University, Shanghai 200240, China \\
$^{3}$ Shanghai Key Laboratory for Particle Physics and Cosmology, Shanghai 200240, China \\
$^{4}$ Waterloo Center for Astrophysics, University of Waterloo, Waterloo, ON N2L 3G1, Canada \\
$^{5}$ Department of Physics and Astronomy, University of Waterloo, Waterloo, ON N2L 3G1, Canada \\
$^{6}$ IPAC, California Institute of Technology, Pasadena, CA 91125, USA\\
}

\begin{document}
\label{firstpage} \pagerange{\pageref{firstpage}--\pageref{lastpage}}
\maketitle
\begin{abstract}

The High Latitude Spectroscopic Survey (HLSS) is the reference baseline spectroscopic survey for NASA's Nancy Grace Roman space telescope, measuring redshifts of $\sim 10$M H$\alpha$ emission line galaxies over a $2000$ deg$^2$ footprint at $z=1-2$. In this work, we use a realistic Roman galaxy mock catalogue to explore optimal phenomenological modeling of the measured power spectrum. We consider two methods for modeling the redshift-space distortions (Kaiser squashing and another with a window function on $\beta$ that selects out the coherent radial infall pairwise velocities, $\Mcur_A$ and $\Mcur_B$, respectively), two models for the nonlinear impact of baryons that smears the BAO signal (a fixed ratio between the smearing scales in the perpendicular and parallel dimensions and another where these smearing scales are kept as a free parameters, P$_{dw}(k|k_*)$ and P$_{dw}(k|\Sigma_\perp,\Sigma_\parallel)$, respectively), and two analytical emulations of nonlinear growth (one employing the halo model and another formulated from simulated galaxy clustering of a semi-analytical model, $\Fcur_{HM}$ and $\Fcur_{SAM}$, respectively). We find that the best model combination employing $\Fcur_{HM}$ is $P_{dw}(k|k_*)*\Fcur_{HM}*\Mcur_B$, while the best combination employing $\Fcur_{SAM}$ is $P_{dw}(k|k_*)*\Fcur_{SAM}*\Mcur_B$, which leads to unbiased measurements of cosmological parameters. We compare these to the Effective Field Theory of Large-Scale Structure perturbation theory model $P_{EFT}(k|\Theta)$, and find that our simple phenomenological models are comparable across the entire redshift range for $k_{max}=0.25$ and $0.3$ $h$/Mpc. We expect the tools that we have developed to be useful in probing dark energy and testing gravity using Roman in an accurate and robust manner.

\end{abstract}

\section{Introduction}\label{sec:Intro}

Next generation wide-field galaxy redshift surveys, such as those from the Dark Energy Spectroscopic Instrument (DESI; \citealt{DESI16}), ESA's Euclid mission \citep{Laureijs11}, and NASA's Nancy Grace Roman Space Telescope \citep{Spergel15}, will measure the 3D distribution of galaxies, collecting more than $10$ times the amount of redshift data currently available. The effective and efficient exploration of the viable cosmological and dark energy parameter space will be important in maximizing the science from Roman.\par

In this paper, we measure the galaxy clustering signal in Fourier-space, using a realistic Roman galaxy mock catalog \citep{Zhai21}\footnote{https://www.ipac.caltech.edu/doi/irsa/10.26131/IRSA546} between $1.0 \le z \le 2.0$, and use different methods to modify the linear predictions of CDM for nonlinear evolution of the dark matter halos and baryonic effects so as to recover the input cosmology, in order to identify a simple phenomenological non-linear power spectrum model that can serve as starting point for more detailed and nuanced explorations of the galaxy clustering signal.

Galaxy redshift surveys such as the Sloan Digital Sky Survey (SDSS; \citealt{Eisenstein05,Anderson12,Anderson14,Ross15,Beutler17}), 2dF Galaxy Redshift Survey (2dFGRS; \citealt{Percival01,Cole05}),  WiggleZ \citep{Blake11,Kazin14}, and 6dF Galaxy Survey (6dFGS; \citealt{Beutler11}) have enabled measurements of the baryon acoustic oscillation (BAO) signal at various redshifts through the observation of galaxy clustering. Additional cosmological information can be gained through the measurement of $f_g\sigma_8$, the growth rate of large-scale structure multiplied by the amplitude of matter fluctuations windowed by a top-hat function of $8$ Mpc/$h$ (\citealt{Blake13,Reid14,Alam17,Zhai19b,Lange22}). These two statistics (BAO and $f\sigma_8$) provide invaluable information about the evolution of the universe and the nature of cosmic acceleration \citep{Guzzo08,Wang08a}, and have so far been consistent with the Planck flat $\Lambda$CDM model \citep{Planck16}. However, there are tensions in the results from current data \citep{Reid14,Zhai2022,Lange22,Yuan2022}. Therefore, it is important to extend the current data set, as well take a closer look at how we model the non-linear clustering signal.

The usual practice for modeling galaxy clustering down into the non-linear regime is to perform a cosmological N-body simulation, extract a halo catalog, and assign galaxies to the host halos with an occupation model (Halo Model; \citealt{Cooray02}). The challenge with the halo model is that you are limited to the cosmology of the N-body simulation. If you wish to have cosmological parameters free you must perform many simulations and employ some technique to interpolate between the cosmologies, i.e., use an emulator to sample the response field (see \citealt{Zhai19b}). Another challenge is that the modeling of the dark matter halo-galaxy connection \citep{Wechsler18} can impact clustering in the non-linear regime, making the emulator specific to a particular selections of galaxies and/or a certain range of redshifts.

We explore corrections to linear cold dark matter (CDM) theory, as predicated using the forward modelling code package {\sc camb}\footnote{\url{CAMB.info}} \citep{Lewis00,Lewis02,Howlett12}, that account for non-linear baryonic and dark matter halo growth effects. For the smearing out of the BAO peak due to baryonic effects, we explore two methods to 'de-wiggling' the BAO signal, one with a single transition parameter $k_*$ \citep{Wang_2013}, that fixes the ratio between the parallel and perpendicular transition scale at $k^\parallel_*=[1+f_g^*(z)]^{-1}k^\perp_*$ where $k^\perp_*=k_*$ (we note is consistent with the method of IR-resummation used to correct Eularian perturbative methods), and another where the parallel and perpendicular transition parameters are left as independent free variables \citep{Beutler17}, $\Sigma_\perp=G(z)/k^\perp_*$ and $\Sigma_\parallel=G(z)/k^\parallel_*$, respectively. For non-linear structure growth, we explore two analytical methods, one which incorporates the halo model ({\sc HaloModel} or HM; \citealt{Mead_2015}) and one with a fitting formula to a galaxy semi-analytical model (SAM) (\citealt{Cole05}), including an analysis on incorporating an additional term for higher $k$'s \citep{Sanchez08}. 

In this work, we use a realistic galaxy mock for the High Latitude Spectroscopic Survey (HLSS), the reference baseline survey for NASA's Nancy Grace Roman space telescope. It covers $2000$ deg$^2$ over the wavelength range of $1-1.93\mu$m (H$\alpha$ redshift range $\sim$1-2), with a depth corresponding to the H$\alpha$ line flux of $10^{-16}$ergs/s/cm$^2$ (6.5$\sigma$), see \citet{Wang22}.
Our results should be qualitatively applicable to other galaxy redshift surveys as well.

This paper is organized as follows: $\S$2 we discuss the power spectrum measurement and modeling techniques used in the simulated observation of our galaxy lightcone mock; $\S$3 we present an analysis of the best modeling techniques and their ability to recover the input cosmology; and in $\S$4 we summarize our findings and point to future applications of the results.

\section{Methodology}\label{sec:Method}
To determine the applicability of the various non-linear corrections to linear theory mentioned in the Introduction, we will use a realistic Roman galaxy mock to measure the galaxy power spectrum monopole and quadrupole moments, $P_0(k)$ and $P_2(k)$, respectively. This measurement will be performed following the standard FKP methodology \citep{Feldman_1994} modified for line-of-sight dependence \citep{Bianchi15} with jackknife covariance estimations. This will require us to model, in addition to the physical non-linear effects, the geometry of the survey with window functions and the redshift-space distortions (RSD). We will then employ a Monte-Carlo Markov Chain (MCMC) statistical analysis to perform a recovery test of the input cosmological parameters. 

\subsection{Mock lightcone galaxy catalog}\label{sec:Mock}

We use the $2000$ deg$^2$ lightcone mock \citep{Zhai21} constructed using the {\sc galacticus} \citep{Benson_2012} semi-analytical model (SAM) and the dark matter {\sc unit} \citep{Chuang_2019} simulation, with cosmological parameters $\Psi=[h,\Omega_b,\Omega_m,\Omega_c,\sigma_8,n_s,A_s]=[0.6774,0.0462,0.3089,0.2627,0.8147,0.9667,
2.06\times10^{-9}]$, where $A_s$ is the amplitude of the primordial matter power spectrum, which correspond to Planck 2015 cosmology model \citep{Planck16}.
The emission line luminosity of the galaxies is computed using the {\sc cloudy} photoionization code \citep{Ferland13}. A full description of the technique can be found in \cite{Merson18}. 

This galaxy mock simulates the galaxy redshift catalog expected from
the Roman High Latitude Spectroscopic Survey (HLSS), where H$\alpha$ galaxies will be observed over an area of $\sim2000$ deg$^2$ mapping out the 3D distribution of $\sim10$ million galaxies at $1.0<z<2.0$. We explore this distribution in redshift slices $z=[(1.0,1.2),(1.2,1.4),(1.4,1.6),(1.6,2.0)]$. Two different dust models where used to calibrate the SAM to replicate either the H$\alpha$ luminosity function observed in the ground-based narrow-band High-z Emission Line Survey (HiZELS; \citealt{Sobral_2012})
or the H$\alpha$ number counts observed in the Hubble Space Telescope (HST) Wide-Field Camera 3 (WFC3) Infrared Spectroscopic Parallel Survey (WISPS; \citealt{Mehta_2015}). In this work, we focus on the dust model calibrated to HiZELS. Note that the observational systematic effects have not been added to the galaxy mock, since they are not yet quantitatively modeled.

\subsection{Power Spectrum Measurement}\label{sec:Pmeasure}

The power spectrum $P(\mathbf{k})$ measures the power of fluctuations in the matter field with wavelengths $\lambda$ denoted by the wavenumber $k=2\pi/\lambda$ ($\mathbf{k}$ is the wavevector which specifies the wavenumber $k$ and the angle relative to the line-of-sight $\theta$ as $\mu=\mathrm{cos}\theta$); it is the Fourier transform of the two point correlation function $\xi(\mathbf{r})$, 
If we define the cosmological overdensity field as $\delta(\mathbf{x})\equiv\rho(\mathbf{x})/\bar{\rho}-1$, where $\rho$ is the matter density and $\bar{\rho}$ is its mean value, then the matter power spectrum is
\begin{equation}
\left< \delta_k \delta_{k'} \right > \equiv (2 \pi )^3 P(\mathbf{k}) \delta^D (\mathbf{k}+\mathbf{k}^{'} ),
\label{eq:Pdef}
\end{equation}
where $\delta^D$ is the Dirac delta function and $\delta_k \equiv \int \mathrm{d}^3 \mathbf{x} \delta(\mathbf{x})\mathrm{exp}(-i \mathbf{k} \cdot \mathbf{x})$ is the Fourier transform of the overdensity $\delta(\mathbf{x})$.\par

We employ the {\sc nbodykit}\footnote{\url{nbodykit.readthedocs.io}} python package, which follows the FKP weighting scheme \blue{for varying line-of-sight \citep{Bianchi15}}, to measure $P(\mathbf{k})$. Since {\sc nbodykit} models $P_{noise}$ as well we will not need to include this in the power spectrum models. The outputs from this measurement will, therefore, have the shot noise removed but will require our theoretical model to be convolved with a window function. 

\begin{figure}
\includegraphics[width=\columnwidth]{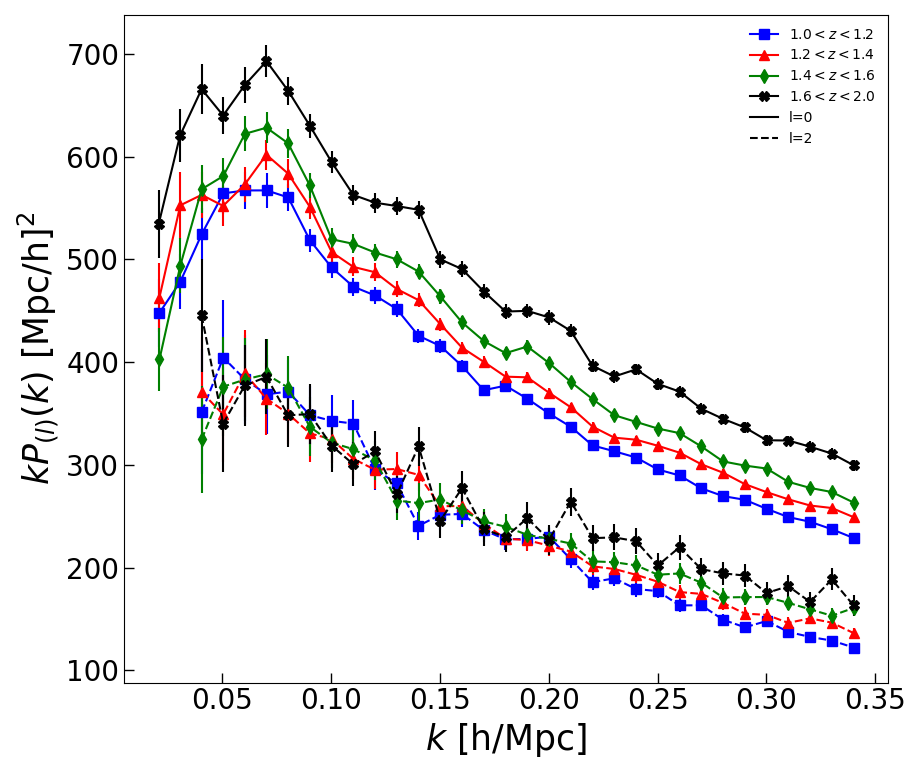}
\caption{Power spectrum multipole measurement scaled by wavenumber ($kP_l$) of the Roman HLSS $2000$ deg$^2$ lightcone mock in 4 redshift slices, $(1.0,1.2),(1.2,1.4),(1.4,1.6),(1.6,2.0)$, denoted by square, triangle, diamond, and cross, respectively. The monopole moments ($P_0$) are shown in solid lines (with shot noise removed) and the quadrupole moments ($P_2$) are shown in dashed lines.}
\label{fig:Pz}
\end{figure}

For this analysis, we model the measurement of the power spectrum monopole moment for the range $0.02 \le k \le k_{max}$ $h\,$Mpc$^{-1}$ and the quadrupole moment for the range $0.04 \le k \le k_{max}$ $h\,$Mpc$^{-1}$, per \citet{Marin16}, with bin-width $\Delta k=0.01$$h\,$Mpc$^{-1}$ for galaxies with an H$\alpha$ flux higher than $10^{-16}$ ergs s$^{-1}$ cm$^{-2}$ for $k_{max}=[0.25,0.3,0.35]$. We perform these measurements for the redshift slices $z=[(1.0,1.2),(1.2,1.4),(1.4,1.6),(1.6,2.0)]$. These measurements are shown in Fig.~\ref{fig:Pz} where the monopole already has the shot noise removed.

\subsubsection{Covariance Matrix}\label{sec:cov}
In order to perform the model likelihood analysis, we need a covariance matrix for the observations. Typically, this is computed using a suite of mock catalogs with the same galaxy clustering statistics as the observational data set.
For instance, \citet{Zhai21} used EZmocks \citep{Chuang_2014} to construct their covariance matrices. 

For convenience, we have chosen to construct our covariance matrix $C$ from jackknife samples of the lightcone mock for each redshift slice. For the data vector $\mathbf{P}=\{P_0,P_2\}$, this is as follows:
\begin{equation}
C(k_1,k_2)=\frac{n-1}{n}\sum_{i=1}^n (\mathbf{P}(k_1)_i-\bar{\mathbf{P}}(k_1))(\mathbf{P}(k_2)_i-\bar{\mathbf{P}}(k_2)),
\label{eq:cov}
\end{equation}
where $n$ is the number of jackknife subsamples, where some portion of the survey is removed and the statistics recomputed over the entire footprint with the sub-sample removed, and $\bar{\mathbf{P}}(k)$ is the average of $\mathbf{P}(k)$ over all the subsamples. We chose to divide the survey footprint into $n=400$ subsamples. 

The problem with the jackknife method is that we are subsampling the single observation that has been made and there is a limit to how often this can be done. Therefore, our approximate covariance matrix will be noisy and how noisy depends on how many subsamples we produce. When the inverse of this noisy matrix is performed, to find the precision matrix $C^{-1}$, our results will be biased. We can correct for the bias in our inversion with either the Hartlap correction \citep{Hartlap_2007} or through a Gaussian smoothing technique \citep{Mandelbaum_2013}, where $C$ is convolved with a 2D Gaussian kernel with width $\sigma$ to remove the noise prior to inversion. We choose to use the Gaussian smoothing technique with a smoothing kernel width $\sigma=1.00$. We find, as have others \citep{Mandelbaum_2013,Lange22,StoreyFisher22}, that this choice gives consistent result as the Hartlap correction.

\subsection{Power Spectrum Modeling}\label{sec:model}

As the universe evolves, prior to recombination and the last scattering of the cosmic microwave background (CMB), there are acoustic oscillations in the baryon-photon fluid that couple gravitationally to the dark matter, which itself is undergoing a mixing of modes across small-scales. These baryon acoustic oscillations (BAO) become frozen post recombination, while the power spectrum amplitude grows with redshift $z$ according to the linear growth factor, $G(z)$. The transfer of power across frequencies, from the primordial $P(k)$ to the beginning of linear growth, is captured by the linear transfer function, $T_{lin}(k)$, resulting in the linear matter power spectrum,
\begin{equation}
P_{lin}(k,z|P_o)=G(z)P_ok^{n_s}T^2_{lin}(k).
\label{eq:Plin}
\end{equation}

The behavior of the linear matter power spectrum is that only the amplitude of this signal will change as the universe evolves, the amplitude being a function of $z$ such that $G(0)=1$ and $P_o$ is the amplitude in the present universe, when $z=0$.

For $T_{lin}$ of matter in a universe with baryons, where there are wiggles in the power spectrum that produce the BAO bump in configuration-space, we utilize the {\sc camb} code package to forward model $T_{lin}$ from cosmological parameters, i.e., $\Psi=[h,\Omega_b,\Omega_m,\Omega_c,\sigma_8,n_s,A_s]$. The shape of the linear transfer function $T_{lin}$ will be mostly determined by the fraction of baryons to matter $\Omega_b/\Omega_m$. In the extreme case where there are no baryons ($\Omega_b=0.0$), no wiggles appear in $T_{lin}$.

\subsubsection{Nonlinear smearing of BAO by baryons}\label{sec:Tdw}

Non-linear baryonic effects have a tendency to broaden the BAO bump in configuration space, which is equivalent to smearing out small-scale wiggles from the BAO signal in $P_{lin}(k)$. This effect can be modeled \citep{Wang_2013} by de-wiggling the BAO signal, transitioning at some non-linear scale $k_*$ to a $T_{lin}$ that has no wiggles, $T_{nw}$, as such:
\begin{equation}
\begin{split}
T_{dw}^2(\mathbf{k},z|k_*)=T_{lin}^2(k)e^{-g_\mu k^2/(2k^2_*)}\\+T_{nw}^2(k)[1-e^{-g_\mu k^2/(2k^2_*)}],
\end{split}
\label{eq:Tdw1}
\end{equation}
with $g_\mu$ as 
\begin{equation}
g_\mu=G(z)^2[1-\mu^2+\mu^2(1+f_g^*(z))^2],
\label{eq:gmu}
\end{equation}
where $f_g^*$ should be equal to the linear growth rate $f_g$ but could be different, and $\mu=\mathrm{cos}(\theta)$, $\theta$ being the angle of $k$ relative to the line-of-sight (LoS). The $e^{-g_\mu k^2/(2k_*^2)}$ term is derived in \cite{Eisenstein_2007}, and captures the effects of nonlinear structure formation on the signature of acoustic oscillations in the late-time galaxy distribution \citep{Angulo08}. The inclusion of the $\mu$-dependence in the $g_\mu$ term, given in Eq.~(\ref{eq:gmu}), accounts for the additional damping along the LoS due to redshift-space distortions (RSD).

\cite{Eisenstein_1998} found a simple fitting formula to this 'non-wiggled' transfer function $T_{nw}$ as
\begin{equation}
\begin{split}
T_{nw}(q)=\frac{L}{L+Cq^2},\\
L(q)=\mathrm{ln}(2e+1.8q),\\
C(q)=14.2+\frac{731}{1+62.5q},\\
q=\frac{k(\mathrm{T}_{CMB}/2.7)^2}{h\Gamma},\\
\Gamma=\Omega_m h
\end{split}
\label{eqn:Tnw}
\end{equation}
where $\mathrm{T}_{CMB}=2.726$ is the CMB temperature at $z=0$. Note that
the last equation in Eqs.(\ref{eqn:Tnw}) assumes zero baryon density, and hence no wiggles. However, there are baryons present and the presence of baryons suppresses power on the broadband, which means we not only need to remove the wiggles as detailed in Eq.~(\ref{eqn:Tnw}) we also need to include the baryonic suppression. This is accomplished by replacing $\Gamma$ with $\Gamma_{eff}$,
\begin{equation}
\begin{split}
\Gamma_{eff}=\Omega_m h \left [\alpha_\Gamma+\frac{1+\alpha_\Gamma}{1+(0.43ks)^4}\right ],\\
\alpha_\Gamma=1-0.328ln(431\Omega_mh^2)\frac{\Omega_b}{\Omega_m}\\
+0.38ln(22.3\Omega_mh^2)\left (\frac{\Omega_b}{\Omega_m}\right )^2,
\end{split}
\label{eqn:GammaEff}
\end{equation}
where
\begin{equation}
s=\frac{44.5ln(9.83/\Omega_mh^2)}{\sqrt{1+10(\Omega_bh^2)^{3/4}}} \mathrm{Mpc}
\label{eqn:s}
\end{equation}
approximates the sound horizon. Therefore, to properly model the effect of non-linear baryon evolution we will use a 'de-wiggled' transfer function $T_{dw}$ as described in Eq.~(\ref{eq:Tdw1}), which transitions from $T_{lin}$ in Eq.~(\ref{eq:Plin}) to $T_{nw}$ of Eq.~(\ref{eqn:Tnw}) at $k>k_*$ with $\Gamma$ replaced by $\Gamma_{eff}$ in Eq.~(\ref{eqn:GammaEff}) to account for baryonic suppression.\par

We can also think of this `de-wiggling' as the addition of a BAO transfer function to the zero-baryon transfer function, where the BAO damps away at $k>k_*$. We see more clearly if we define $T_{BAO}^2(k)=T_{lin}^2(k)-T_{nw}^2(k)$, which makes Eq.~(\ref{eq:Tdw1}) as follows:
\begin{equation}
\begin{split}
T_{dw}^2(\mathbf{k},z|k_*)=T_{nw}^2(k)+T_{BAO}^2(k)e^{-g_\mu k^2/(2k^2_*)}.
\end{split}
\end{equation}

Although $f_g^*$ is allowed to be a free parameter, it cannot be measured from actual data due to parameter degeneracies, so it's usually fixed at the growth rate factor $f_g(z)$, thus effectively fixing the ratio of damping scales in the perpendicular and parallel $k$-dimensions.
To mitigate this, we could go back to the more basic equation:
\begin{equation}
\begin{split}
T_{dw}^2(\mathbf{k}|\Sigma_\perp,\Sigma_\parallel)=T_{nw}^2(k)+T_{BAO}^2(k)e^{-[(1-\mu^2)\Sigma_\perp^2+\mu^2\Sigma_\parallel^2]k^2/2},
\end{split}
\label{eq:Tdw2}
\end{equation}
where $G(z)/k^\perp_*$ is replaced by $\Sigma_\perp$ and $G(z)/k^\parallel_*$ by $\Sigma_\parallel$.\par

The `de-wiggled' power spectrum is then
\begin{equation}
P_{dw}(\mathbf{k},z|P_o,\psi)=G(z)P_ok^{n_s}T^2_{dw}(\mathbf{k},z|\psi),
\label{eq:Pdw}
\end{equation}
where $\psi=k_*$ when there is a fixed ratio between parallel and perpendicular dimensions, or $\psi=\Sigma_\perp,\Sigma_\parallel$ 
when each dimension is free. Note that the $k_*$ technique effectively fixes $\Sigma_\parallel/\Sigma_\perp=1+f_g^*(z)$. Also, for the [$\Sigma_\perp,\Sigma_\parallel$] technique, $T_{dw}$ does not explicitly depend on $z$.\par

\subsubsection{Nonlinear Structure Growth}\label{sec:NLgrowth}

The growth of the matter power spectrum post-recombination undergoes further transfer across frequencies as gravity mixes the small-scale modes. We do not detail the non-linear growth according to CDM as a function of $z$ but rather employ an analytical correction to the linear theory found when the linear power spectrum was corrected to the non-linear clustering result of N-body simulations: either the non-linear matter power spectrum inferred from clustering according to the halo model ({\sc HMcode} or $\Fcur_{HM}$; \citealt{Mead_2015}) or the non-linear galaxy power spectrum found using a galaxy catalog constructed with a semi-analytic model (SAM) ($\Fcur_{SAM}$; \citealt{Cole05}).

The non-linear corrections of \cite{Mead_2015} uses the halo model directly in the modelling framework as opposed to the {\sc halofits} method of \citep{Smith_2003,Takahashi_2012} that employs an empirical fitting algorithm to N-body simulation clustering statistics but does not directly use the halo model equations in their formula. Additionally, \cite{Mead_2015} incorporate parameters to account for the impact on the dark matter from baryonic feedback. We performed a few minimization tests leaving the baryonic feedback parameters free and found best fits similar to the default values as set in {\sc camb}. Since we also found little dependence on the particular choice of fixed values we left them fixed at default. \cite{Joachimi21} found results similar to these defaults in their analysis of the KiDS-1000 joint galaxy clustering and weak lensing analysis.

The nonlinear correction to $P_{lin}$ according to {\sc HMcode} for our chosen cosmology is simply the ratio:
\begin{equation}
\Fcur_{HM}(k,z)=\frac{P_{nl}(k,z)}{P_{lin}(k,z)}=\frac{T^2_{nl}(k,z)}{T^2_{lin}(k)},
\label{eq:Fhf}
\end{equation}
where the non-linear $P_{nl}$ and linear $P_{lin}$ power spectrum has been fully predicted to $z$ with the {\sc camb} code, the ratio of which becomes the ratio of the linear transfer functions squared when $G(z)P_ok^{n_s}$ drop out. Multiplying $P_{lin}$ by $\Fcur_{HM}$ then gives us the non-linear matter power spectrum with baryonic feedback impacting the distribution of matter. \cite{Joachimi21} found {\sc HMcode} to be comparable to the state-of-the-art emulator {\sc CosmicEmu} \citep{Heitmann14}.

Another method we use to incorporate non-linear growth into our model is similar to the {\sc halofits} technique, in that a fitting formula is used to find the correction to linear theory, but fits to a galaxy catalog rather than the matter density field via halo clustering. \cite{Cole05} found that when comparing linear matter theory to the galaxy catalog output by a SAM \citep{Benson00} (an earlier version of the {\sc galacticus} code that is used to create our mock galaxy lightcone) where the halo merger trees have been used to paint galaxies onto the halos, the following function can be used:
\begin{equation}
\Fcur_{SAM}(k|Q_1,Q_2(,Q_3))=\frac{1+Q_1k^2}{1+Q_2k(+Q_3k^2)},
\label{eq:Fsam}
\end{equation}
which when multiplied by \boldblue{}{$b_g^2P_{lin}$} produces the nonlinear galaxy power spectrum. The $Q_1$ term has more impact on small-scales while $Q_2$ has more of an impact on large-scales. The additional term in parenthesis $Q_3$ was proposed by \cite{Sanchez08} to improve modeling across larger $k$-values. We will explore the need for $Q_3$ in this work. 

\cite{Cole05} employed this non-linear correction with the linear prediction of {\sc CAMB} to look for any non-linearity and scale-dependent bias present in their new estimator used to measure the power spectrum of the 2dF Galaxy Redshift Survey. They compared using $\Fcur_{SAM}$ to predictions of {\sc halofits} (see their paper $\S$7.2) for different cosmologies and found it to be fairly robust, in that the general trend of non-linearity in the matter power spectrum are well represented by the $\Fcur_{SAM}$ correction even when the cosmology is varied. That being said, the benefit of using $\Fcur_{SAM}$ over {\sc halofits}, or more importantly for this work $\Fcur_{HM}$, is that $\Fcur_{SAM}$ also includes the non-linearity of the galaxy population and deviations from the linear galaxy bias $b_g$, which is difficult to predict.\par

For the purposes of our work, we consider $\Fcur_{SAM}$ as an emulator that has been constructed from a realistic galaxy biasing scheme, the semi-analytic model, and has been used to test the systematics of an estimator used to measure actual data. Additionally, it has been shown to be fairly robust to changes in cosmology. We should note that $\Fcur_{SAM}$ was not tested for robustness with regards to changes in $z$. Something we will test in this work.

We then have one non-linear correction to the linear matter power spectrum and one non-linear correction to the linear galaxy power spectrum, $\Fcur_{HM}$ and $\Fcur_{SAM}$, respectively. The non-linear galaxy power spectrum in real-space is then either
\begin{equation}
P_{nl,g}(k,z)=b_g^2\Fcur_{HM}(k,z)P_{lin,m}(k,z),
\label{eq:PnlHM}
\end{equation}
or
\begin{equation}
P_{nl,g}(k)=\Fcur_{SAM}(k)P_{lin,g}(k,z),
\label{eq:PnlSAM}
\end{equation}
where $P_{lin,g}=b_g^2P_{lin,m}$ and $P_{lin,m}$ is either the linear matter power spectrum detailed in Eq.~(\ref{eq:Plin}) or the 'de-wiggled' linear matter power spectrum detailed in Eq.~(\ref{eq:Pdw}). \par

It should be noted that the only difference between Eq.~(\ref{eq:PnlHM}) and Eq.~(\ref{eq:PnlSAM}) is the use of either $\Fcur_{HM}$ or $\Fcur_{SAM}$. The reason we present them as two different equations is to emphasize what these non-linear corrections are actually correcting. The correction according to {\sc HMcode}, $\Fcur_{HM}$, is producing the non-linear matter power spectrum through a phenomenological correction to the linear matter power spectrum, which is then used to produce the non-linear galaxy power spectrum assuming a linear galaxy bias, $b_g$. This is different than the phenomenological correction found through analysis of galaxy clustering, $\Fcur_{SAM}$, which produces the nonlinear galaxy clustering signal in Fourier-space by serving as a prefactor to the linear galaxy power spectrum, itself a result of assuming a linear galaxy bias $b_g$. Therefore, even though these equations look exactly the same minus the different correction factors, $\Fcur_{SAM}$ accounts for additional non-linear and scale-dependent bias introduced by the galaxy population whereas $\Fcur_{HM}$ does not.

Since our galaxy mock catalog was built using a SAM (see $\S$\ref{sec:Mock}), the naive expectation is that $\Fcur_{SAM}$ should be the correct model versus using $\Fcur_{HM}$. That being said, we are observing only galaxies above a particular $H\alpha$ flux cut and between redshifts $1.0<z<2.0$, therefore it may be that the non-linear galaxy clustering signal is not as nuanced as at lower redshifts and either method is applicable. This is one of the main things we are looking to explore in this work. How important are the non-linear corrections in the Roman universe?

\subsubsection{Model Selection Choices}\label{sec:MODchoice}

We have discussed $2$ different ways to model the non-linear corrections to linear theory ($\Fcur_{HM}$ and $\Fcur_{SAM}$) and $2$ different methods to de-wiggle the BAO signal due to non-linear growth ($P_{dw}$ in Eq.~(\ref{eq:Pdw}) with either $\psi=k_*$ or $\psi=\Sigma_\perp,\Sigma_\parallel$). Given that $\Fcur_{HM}$ and $\Fcur_{SAM}$ have been constructed as corrections to the $3D$ power spectrum $P(k)$, it is not known if replacing $P_{lin}$ in Eqs~(\ref{eq:PnlHM}) and ~(\ref{eq:PnlSAM}) with $P_{dw}$ will improve the model or not. $\Fcur_{SAM}$ was shown in \cite{Cole05} to reproduce real- and redshift-space clustering of $P(k)$, but there may be trends with $\mu$ that are not captured after the $3D$ integration, while for $\Fcur_{HM}$ was constructed in real-space and so might benefit more from the inclusion of $P_{dw}$. We will explore all $3$ choices, $P_{lin}(k)$ (Eq.~(\ref{eq:Plin})) or with $P_{dw}(k|k_*)$ or $P_{dw}(k|\Sigma_\perp,\Sigma_\parallel)$ (Eq.~(\ref{eq:Pdw})), for the nonlinear correction $\Fcur_{HM}$ (Eq.~(\ref{eq:Fhf})) and for $\Fcur_{SAM}$ with and without $Q_3$ (Eq.~(\ref{eq:Fsam})).\par

\subsection{Redshift Space Distortions}\label{sec:RSD}
In real-space, where the universe is isotropic, there is no preferred LoS. The clustering of galaxies is spherical. However, we observe galaxies from redshift surveys in redshift-space and must account for RSDs, where there are additional redshifts due to the peculiar motions of galaxies within the Hubble flow (see \cite{Hamilton_1998} for review).

On larger scales, there is a squashing of the two-point correlation function (2PCF) as galaxies fall into overdensities causing an additional red- or blue-shift along the LoS. The squashing effect of the RSD is captured by the Kaiser \citep{Kaiser_1987} factor $(1+\beta\mu^2)^2$ with the anisotropic parameter $\beta=f_g/b_g$ detailing deviations from sphericity, where $f_g$ is the linear growth rate, and $b_g$ is the tracer bias. The linear growth rate $f_g$ is the change in the linear growth factor $G$ with scale factor $a$, i.e.,
\begin{equation}
f_g(a)=\frac{d\mathrm{ln}G}{d\mathrm{ln}a},
\label{eq:f}
\end{equation}
which can be approximated as the cosmic matter density 
$\Omega_m(a) \equiv 8\pi G\rho_m(a)/(3H^2(z))$ raised to the growth index, $\gamma$ (\citealt{Wang_1998}), that is a prediction of the cosmological model ($\Lambda CDM$ predicts $\gamma \simeq 0.55$, see \citealt{Lue04}).

On smaller scales, galaxy peculiar velocities
result in elongations of the 2PCF at small perpendicular separations. Since these elongations are always along the LoS, it appears as if they are always pointing back to the observer and are so denoted as the Finger-of-God (FoG) effect. The FoG is most often analytically incorporated to the galaxy clustering by either an exponential distribution or a Gaussian distribution, the Fourier transform of which produces either a Lorentz damping or Gaussian damping term \citep{Percival_2009}.\par

We model the RSD in two ways. The first has the standard Kaiser term for large-scale squashing and an exponential distribution for the small-scale FoG, which becomes a Lorentzian damping term in Fourier space. We refer to this method for modeling the RSD as $\Mcur_A$ and is:
\begin{equation}
\Mcur_A(\mathbf{k}|\beta,\sigma_{r,v})=\frac{(1+\beta\mu^2)^2}{1+\frac{1}{2}(k\mu\sigma_{r,v})^2}
\label{eq:ModA}
\end{equation}
where $\sigma_{r,v}=\sigma_v/[H(z)a(z)]$, with $\sigma_v$ being the pairwise velocity dispersion in [km/s] of the galaxies.

The second method for modeling the RSD takes into account that the standard Kaiser term is derived with a curl-free assumption about the pairwise velocity and that an actual measurement of the RSD will include some contribution from velocities that have a curl. \cite{Zhang_2013} decomposed the peculiar velocity into three components: an irrotational component correlated(uncorrelated) with the underlying density field, \textbf{v}$_\delta$(\textbf{v}$_S$), and a rotational component \textbf{v}$_B$. They found that selecting the cosmological information from the irrotational correlated velocity term \textbf{v}$_\delta$ is equivalent to applying a window function ($\tilde{W}$) to the $\beta$ parameter in the Kaiser term. \cite{Zheng13} found that with this windowed $\beta$ modification a Gaussian distribution for the FoG was best able to recover the expected cosmology. Thus we have our second model for the RSD
\begin{equation}
\Mcur_B(\mathbf{k},z|\beta,\sigma_{r,v},\Delta\alpha)=\left(1+\beta\tilde{W(k,z)}\mu^2\right)^2\rm{exp}\left[-\frac{(k\mu\sigma_{r,v})^2}{2}\right]
\label{eq:ModB}
\end{equation}
where
\begin{equation}
\tilde{W}(k,z)=\frac{1}{1+\Delta\alpha(z)\Delta^2(k,z)}
\label{eq:Wbeta}
\end{equation}
with $\Delta\alpha(z)$ being a free parameter to be determined by observational data and the dimensionless power spectrum $\Delta^2(k,z)=k^3P_{lin}/(2\pi^2)$. 

In spite of the apparent difference between Lorentzian and Gaussian modeling for the FoG, the primary difference between $\Mcur_A$ and $\Mcur_B$ is having a window function or not on $\beta$. In $\Mcur_B$, it is the velocity component produced by applying the window function that is Gaussian, not a Gaussian applied to a non-windowed $\beta$ that displays more non-Gaussian characteristics. Peculiar motions on small-scales that have a superposition of in-falling and orbiting galaxies are non-Gaussian, perhaps being better fit by a Lorentzian profile, whereas the coherent in-falling galaxies, which are isolated by applying the window function, are better fit by a Gaussian profile. 

We then have two ways to transpose the real-space clustering signal into redshift-space, either
\begin{equation}
P'_{nl,g}(\mathbf{k},z)=\Mcur_A(\mathbf{k}) P_{nl,g}(k,z)
\label{eq:PnlModA}
\end{equation}
or
\begin{equation}
P'_{nl,g}(\mathbf{k},z)=\Mcur_B(\mathbf{k},z) P_{nl,g}(k,z),
\label{eq:PnlModB}
\end{equation}
where $'$ denotes the Fourier galaxy clustering signal in redshift-space and $P_{nl,g}$ is the Fourier non-linear galaxy clustering in real-space as described by either Eq.~(\ref{eq:PnlSAM}) or Eq.~(\ref{eq:PnlHM}), resulting in a total of $6$ choices for $\Fcur_{HM}$ and $12$ choices for $\Fcur_{SAM}$, which we explore in Tables~\ref{table:ModCompHM} and \ref{table:ModCompSAM}, respectively.\par

From this point moving forward, we will forgo including the redshift $z$ in the model functions and note that we will perform the fits to each redshift slice assuming a fixed redshift at the center of the redshift slice bin, i.e., for redshift slices $z=[(1.0,1.2), (1.2,1.4), (1.4,1.6), (1.6,2.0)]$ we use $z_{cen}=[1.1,1.3,1.5,1.8]$ in the respective power spectrum model functions.\par

\subsection{Redshift-Space Multipoles}
In redshift-space, where the clustering is no longer isotropic, additional information can be gained by measuring the higher order multipoles of the power spectrum
\begin{equation}
P_l(k)=\frac{2l+1}{2}\int^1_{-1}P'(\mathbf{k})\lcur_l(\mu)d\mu,
\label{eq:Pmlt}
\end{equation}
such that the sum of the multipoles produces the total power spectrum, i.e.,
\begin{equation}
P'(\mathbf{k})=\sum^\infty_{l=0,2,4,\cdots}P_l(k)\lcur_l(\mu),
\label{eq:Psum}
\end{equation}
where $\lcur_l$ is the Legendre polynomial. For the linear power spectrum, the three non-vanishing multipoles are the monopole ($l=0)$, quadrupole ($l=2$), and hexadecapole ($l=4$). As discussed in $\S$\ref{sec:Pmeasure}, we will model the monopole and quadrupole moments, $P_0$ and $P_2$, respectively. \par

\subsection{Survey window}\label{sec:wind}

Now with a theoretical model for the galaxy power spectrum, we simply need to convolve $P(k)$ with a window function 
to account for the finite survey volume, as follows:
\begin{equation}
P^c(\mathbf{k})=\int \frac{d^3\mathbf{q}}{(2\pi)^3}P'(\mathbf{q})|W_2(\mathbf{k}-\mathbf{q})|^2.
\label{eq:Pwind}
\end{equation}
For our purpose, it is efficient and sufficient to compute this by performing a convolution with the power spectrum multipoles requiring only 1D FFTs. We will follow the method detailed in \citet{Beutler_2016} that converts the monopole and quadrupole from Fourier into configuration-space, applies a survey mask, and then performs an inverse transformation to the corrected correlation functions to model the window corrected power spectrum multipoles. This technique uses Hankel transformations to perform the FFTs, going from Fourier to configuration-space with
\begin{equation}
\xi_l(s)=\frac{4\pi(-i)^l}{(2\pi)^3}\int dk k^2 P_l(k)j_l(sk),
\label{eq:F2C}
\end{equation}
where $j_l$ is the spherical Bessel function of order $l$.

Specifically, we use the monopole and quadrupole with appropriate window masks
to find the corrected correlation functions \citep{Wilson17},
\ba
\xi_0^c &=& \xi_0W_0^2 + \frac{1}{5}\xi_2W_2^2 +\cdots\\
\label{eq:xiWIND}
\xi_2^c &=& \xi_0 W_2^2+\xi_2 \left [ W_0^2 + \frac{2}{7}W_2^2+\frac{2}{7}W_4^2 \right ]+\cdots
\label{eq:xiWIND2}
\ea
where the window function multipoles can be derived from the random pair distribution as
\begin{equation}
W_l^2(s)\propto RR(s,\mu)\lcur_l(\mu)
\label{eq:windMLT}
\end{equation}
with the normalization that $W_0^2(s\rightarrow 0)=1$.

After applying these window functions to the two-point correlation statistics, we perform an inverse transform to get back the window function corrected power spectrum monopole and quadrupole moments, via
\begin{equation}
P^c_l(k)=4\pi(-i)^l\int dk k^2 \xi^c_l(s)j_l(sk).
\label{eq:C2F}
\end{equation}

\subsection{Scaling parameters}\label{sec:AP}
In constructing a realistic model for the power spectrum in redshift-space, we have assumed a fiducial cosmological model (with parameters $\Psi$). In actual measurements, if the true cosmology is different than the fiducial, there will be a distortion in the 3D comoving coordinates known as the Alcock-Paczynski, or AP, effect \citep{Alcock_1979}. To account for this effect, we introduce two scaling parameters, parallel and perpendicular to the line-of-sight,
\begin{equation}
\alpha_\parallel=\frac{H^{fid}(z)r_s^{fid}(z_d)}{H(z)r_s(z_d)}, \hskip 0.5cm \alpha_\perp=\frac{D_A(z)r_s^{fid}(z_d)}{D_A^{fid}(z)r_s(z_d)},
\label{eq:alpha}
\end{equation}
where $H$ and $D_A$ are the Hubble parameter and angular diameter distance for the fiducial model, and $r_s$ is the sound horizon at the drag epoch, the superscript $fid$ indicating predictions from the fiducial model. These scaling terms will distort the parallel and perpendicular components of the $k$-vector: $\tilde{k}_\parallel=k_\parallel/\alpha_\parallel$ and $\tilde{k}_\perp=k_\perp/\alpha_\perp$. This becomes a distortion in the $k$ and $\mu$ via
\begin{eqnarray}
\tilde{k}=\frac{k}{\alpha_\perp}\left [1+\mu^2\left(\frac{1}{F^2}-1\right)\right]^{1/2},\\
\tilde{\mu}=\frac{\mu}{F}\left [1+\mu^2\left(\frac{1}{F^2}-1\right)\right]^{-1/2},
\label{eqn:APkmu}
\end{eqnarray}
where $F=\alpha_\parallel/\alpha_\perp$. This modification will require the multipole moments be multiplied by $(\alpha_\perp^2\alpha_\parallel)^{-1}$ such that Eq.~(\ref{eq:Pmlt}) becomes
\begin{equation}
P_l(k,z)=\frac{2l+1}{2\alpha_\perp^2\alpha_\parallel}\int^1_{-1}P'(\mathbf{\tilde{k}},z)\lcur_l(\tilde{\mu})d\mu.
\label{eq:PmltAP}
\end{equation}

By incorporating this AP effect into our power spectrum model, we can very simply test our cosmological model through the constraints on $\alpha_\parallel$ and $\alpha_\perp$, the expectation being that if our fiducial cosmology is correct we will find $\alpha_{\parallel,\perp}=1.0$. These terms allow us to constrain cosmology from distortions in the shape of the BAO signal, detailing how the parallel and perpendicular dimensions are altered from an incorrect cosmological prior. 

Similarly, we can use an alpha term to parameterize deviations from the fiducial linear growth parameter \citep{Wang_2013},
\begin{equation}
f_g(z)\sigma_m(z)\equiv f_g(z)G(z)\sqrt{P_o},
\label{eq:growthparam}
\end{equation}
relative to predictions of the fiducial cosmology as \cite{Zhai21}
\begin{align}
\alpha_g \equiv \frac{f_g(z)\sigma_m(z)}{f_{g,fid}(z)\sigma_{m,fid}(z)}= \frac{f_g(z)G(z)\sqrt{P_o}}{f_{g,fid}(z)G_{fid}(z)\sqrt{P_{o,fid}}}\\
= \frac{\beta(z)\sqrt{P_n(z)}}{f_{g,fid}(z)G_{fid}(z)\sqrt{P_{o,fid}}},
\label{eqn:alphag}
\end{align}
where $P_n(z)=P_oG^2(z)b^2(z)$. Compared to the widely used $f_g(z)\sigma_8(z)$ parametrization, $f_g(z)\sigma_m(z)$ has the advantage of having no explicit dependence on the Hubble constant.

\subsection{Parameter Constraints and Model Evaluation}\label{sec:cons}

To test our nonlinear models, we will evaluate how well each model is able to recover the input cosmology through analysis on the constraints of $\alpha_\perp$, $\alpha_\parallel$, and $\alpha_g$. 
We expect $\alpha_{\perp,\parallel,g}=1.0$ within $1\sigma$ if our modeling is sufficiently accurate.

To find the posterior distribution of the modelling parameters, we will perform a likelihood analysis with the Markov Chain Monte Carlo (MCMC) technique, with $\chi^2$ as follows:
\begin{equation}
\chi^2=\sum_{i,j}(\mathbf{P}_{obs,i}-\mathbf{P}_{th,i})C^{-1}_{ij}(\mathbf{P}_{obs,j}-\mathbf{P}_{th,j}),
\label{eq:chi2}
\end{equation}
where the index $i$($j$) indicates the data vector at $k_i$($k_j$), $\mathbf{P}_{obs}$ is from the measurement of the mock galaxy lightcone while $\mathbf{P}_{th}$ is the prediction of the data vector from theory, monopole and quadrupole in Eq.~(\ref{eq:C2F}) respectively, and $C^{-1}$ is the inverse of the covariance matrix found in Eq.~(\ref{eq:cov}).

All models have the set of parameters $[\beta,b_g^2P_o,\sigma_v,\alpha_\perp,\alpha_\parallel]$, with priors $\beta=[0.3,0.85]$, log$_{10}(b_g^2P_o)=[4,8]$, $\sigma_v=[10,700]$, $\alpha_\perp=[0.5,1.5]$, and $\alpha_\parallel=[0.5,1.5]$. The linear growth parameter is a derived quantity according to Eq.~\ref{eqn:alphag}, and so is a combination of the $\beta$ and $b_g^2P_o$ constraints. FoG models, $\Mcur_A$ and $\Mcur_B$ have the parameter $\sigma_v$. If $\Mcur_B$ is used, there is an additional parameter $\Delta\alpha$ with priors $\Delta\alpha=[0,1.6]$. For the single parameter de-wiggle power spectrum the parameter and bounds are $k_*=[0.01,1.5]$. For the two parameter de-wiggle model we have the parameters $[\Sigma_\perp,\Sigma_\parallel]$ both with bounds $[0.5,30.0]$. The {\sc HaloModel} nonlinear correction $\Fcur_{HM}$ does not have any additional parameters since, as discussed in $\S$\ref{sec:NLgrowth}, we fix the baryon feedback parameters to the default value in {\sc camb}. For $\Fcur_{SAM}$ there are three parameters $[Q_1,Q_2,Q_3]$ with bounds $[(0.01,15.0),(2.0,30.0),(0.2,10.0)]$.

We could continue to add parameters in an attempt to fit the data better but this can lead to overfitting, and from a physical perspective, we should adhere to Occam's Razor (also known as the Law of Parsimony; \citealt{Froidmont1649}) and choose the simplest model. We use the Bayesian Information Criteria (BIC) to evaluate model selection
in this context:
\begin{equation}
BIC=\chi^2+m\mathrm{ln}(n),
\label{eq:BIC}
\end{equation}
with $n$ being the number of data points fit and $m$ the number of modelling parameters, which has the same variables as $\chi^2_r$ ($\chi^2$ per degree of freedom) for the special case of Gaussian errors, but is derived from information theory and accomplishes our goal of understanding how well a model fits the data while also penalizing models with more parameters. \par

Perhaps we should then select the model with the lowest $BIC$ value, results presenting in Table \ref{table:ModCompHM} and \ref{table:ModCompSAM}, and call that our best model. However, fitting the data more correctly with fewer parameters does not mean we have the most physically relevant model. For instance, a particular model could chase the data points with smaller errors and miss the data with larger errors, which in this case means fitting the small-scale clustering data but missing the large-scale clustering sensitive to cosmology resides.\par

Since we are more interested in how well a model is able to recover the input cosmology through analysis of the posteriors on $\alpha_{\perp,\parallel,g}$, any model that has $\alpha_{\perp,\parallel,g}$ within $1\sigma$ of unity is a good model.  We can quantify this with the Figure-of-Bias (FoB) value \citep{Eggemeier21}. The FoB for the three alpha parameters is as follows:
\begin{equation}
FoB=\left[ \sum_{i,j}(\bar{\theta}_i-\theta_{fid,i})S^{-1}_{ij}(\bar{\theta}_j-\theta_{fid,j})\right]^{1/2},
\label{eq:FoB}
\end{equation}
where $\bar{\theta}=[\bar{\alpha}_\perp,\bar{\alpha}_\parallel,\bar{\alpha}_g]$ is the average of the posterior distributions of the $\alpha$-terms from the MCMC chains, $\theta_{fid}=[1,1,1]$, and $S_{ij}$ is the covariance matrix of the $[\alpha_\perp,\alpha_\parallel,\alpha_g]$ posterior distribution. Generally speaking, a smaller FoB, results presented in Table \ref{table:FoB}, means better accuracy in recovering the input cosmological model.

That being said, it is possible to be accurate but not precise and the FoB alone will not tell us that, since a lower FoB can be obtained by not having a mean value closer to the fiducial value but by rather increasing $\sigma$. To assess how the areas of the posteriors are different between each model, we employ the Figure-of-Merit (FoM) statistic \citep{Wang08b},
\begin{equation}
FoM=\frac{1}{\sqrt{\mathrm{detS}_{ij}}}.
\label{eq:FoM}
\end{equation}
A larger FoM value indicates a smaller constraint area. While the area of $S_{ij}$ is a contribution to the FoB, as mentioned previously, we cannot disentangle a lower FoB caused by the mean value of the constraint being closer to the fiducial value rather then being caused by a larger contour, larger standard deviation. With the FoM in addition to the FoB, we can consider both values, looking for a lower FoB with a high FoM, to determine which models are not only accurate but also more precise.\par

\subsection{Effective Field Theory of Large-Scale Structure}\label{sec:eft}

The phenomenological nonlinear power spectrum models explored in this work are an attempt to move beyond the linear behavior of structure growth that assumes the amplitude of the matter power spectrum grows uniformly such that the final matter distribution is simply a multiple of the initial distribution and that the galaxy distribution is a multiple of this, which implies the usage of the linear transfer function $T_{lin}$ and the linear galaxy bias $b_g$. In reality, dark matter particles self gravitate forming halos such that modes of the matter distribution grow at different rates on small scales and the biasing between galaxy and the matter distribution has a scale-dependence.\par

The galaxy density field can more accurately be expressed as a function of the matter density that we can then Taylor expand \citep{Fry93}, i.e.,
\begin{equation}
\delta_g(\mathbf{x})=f(\delta(\mathbf{x}))=\sum^\infty_{i=0}\frac{b_i}{i!}\delta^i_m(\mathbf{x}),
\label{eq:deltag}
\end{equation}
such that we can think of $b_g$ as the first-order term in this expansion, which is linear in $\delta_m$. Including the nonlinear bias means to include the higher order terms in this expansion. If we want to account for the nonlinear evolution of the mass density field, given that the evolution of the mass density in large-scale structure is described by fluid equations, we can then employ the Eulerian standard perturbation theory (SPT; \citealt{Bernardeau02}) that solves the continuity and Euler equations, treating the matter as a pressureless fluid, as is done in \cite{Beutler17b}, to expand $\delta_m$ as such,
\begin{equation}
\delta_m(\mathbf{k})=\sum^\infty_{j=1}\delta_m^{(j)}(\mathbf{k}),
\label{eq:deltam}
\end{equation}
where the expansion terms can be expressed according to their Feynman diagrams, i.e., tree-level and loop contributions \citep{Crocce06}.

The nonlinear correction $\Fcur_{HM}$ is then an attempt to include in the model a function that will estimate the higher order nonlinear mass density contributions in Eq.~(\ref{eq:deltam}) ($\delta_m^{(1)}$ being the linear mass density field), while $\Fcur_{SAM}$ is an attempt to do this, as well as account for the convergent value of $i>1$ in the sum of Eq.~(\ref{eq:deltag}), both based on the results of large-scale structure growth simulations, with and without galaxies, $\Fcur_{SAM}$ and $\Fcur_{HM}$, respectively.

These phenomenological nonlinear corrections may be very useful since getting very far beyond 1-loop corrections in SPT requires solving complicated integrals that are computationally expensive and can diverge. That being said, performing a Fourier transform on the integrals that then become simple multiplication in position space can be done with the Fast Fourier Transform method (FFT; \citealt{Schmittfull16}), which makes computing the 1- and 2-loop integrals more convenient allowing for usage of SPT in MCMC likelihood analysis. The FFT method has been employed to compute the 1-loop integrals in the publicly available {\sc fast-pt} code developed by \cite{McEwen16}. 

Even then, 1-loop SPT deviates up to $20\,\%$ for $k \leq 0.2\,h$/Mpc \citep{Scoccimarro04}, the accuracy of the order of the 2-loop contribution is 
$P_{2-loop}/P_{lin}\sim 6\,\%$ at $z=0$ and $k=0.1\,h$/Mpc, while the 3-loop SPT 
shows less agreement than the 1-loop correction even in linear regimes and 
diverges at $k>0.16\,h$/Mpc \citep{Blas14} at low redshifts. SPT can be improved 
by partially resumming the infinite series of higher-order perturbations via 
renormalization perturbation theory (RPT; \citealt{Crocce06}) that employs the 
Zel'dovich approximation \citep{Zel'dovich70} and is based on the Lagrangian 
perturbation theory (LPT; \citealt{Buchert92}), where the displacements of fluid 
elements are treated as dynamical variables \citep{Okamura11}. However, this 
improvement alone is not enough to account for bulk flows that impact the BAO 
signal, where small scales affect large scales, since even with resummation 
(renormalization) PT theory does not converge.

This has typically resulted in two modifications to SPT. First, is treating the matter density field as an effective fluid that has small perturbations and is characterized by a few parameters like an equation of state, a sound speed and a viscosity parameter via the effective field theory (EFT; \citealt{Baumann12,Carrasco12,Senatore14}). The intent here is to smooth out the short-wavelength modes at $k>\Lambda$ through a convolution of a window function with the density field thereby decomposing the field into long- and short-wavelength contributions \citep{Baumann12}, i.e.,
\begin{equation}
\delta(\mathbf{x})=\int d^3\mathbf{x}W_\Lambda(\mathbf{x}-
\mathbf{x}')\delta(\mathbf{x}')+\delta_s\equiv\delta_l
(\mathbf{x})+\delta_s(\mathbf{x}),
\label{eq:deltaEFF}
\end{equation}
that allows for SPT to more accurately model the long-wavelength universe with the addition of counterterms, additional terms that are added to the action to renormalize the theory \citep{Goswami14}, that account for the impact of the short-wavelength universe on the observation. These counterterms result from a modification to the Euler and continuity equation that includes an effective stress tensor that is sourced by the short-modes \citep{Carrasco12}, are on the order of $\eta k^2P_{lin}$, and serve to cancel the divergent terms present in SPT \citep{Pajer13}. So, in short, EFT attempts to isolate the linear universe allowing for SPT to be used on the long-wavelengths with the addition of quantum counterterm corrections from the short-wavelengths that cause the model to converge.

A second modification to SPT, which is required even with the EFT modifications, is performing an infrared (IR) resummation \citep{Senatore15,Blas16,Ivanov18} that attempts to model the behavior of coupled modes at short-wavelengths that result in longer wavelength displacements. This entails the decomposition of the power spectrum into a `wiggly' part and a `non-wiggly' part \citep{Ivanov20},
\begin{equation}
P_{lin}(\mathbf{k}) = P_{nw}(\mathbf{k}) + P_{w}(\mathbf{k}),
\label{eq:PIRdecomp}
\end{equation}
where the IR resummed anisotropic power spectrum at leading order takes the following form,
\begin{equation}
P_{LO}(\mathbf{k})= P_{nw}(\mathbf{k}) + e^{-k^2\Sigma_{tot}^2(\mu)}P_w(\mathbf{k}),
\label{eq:PIRlo}
\end{equation}
which has the same form as our implementation of the `de-wiggled' power spectrum seen in Eq.~(\ref{eq:Tdw2}), with
\begin{equation}
\Sigma_{tot}^2=(1+f\mu^2(2+f))\Sigma^2+f^2\mu^2(\mu^2-1)\delta\Sigma^2,
\label{eq:SIGtot}
\end{equation}
which we see has the exact same form of our `de-wiggled' $k_*$ model (Eq.~(\ref{eq:gmu})) if we ignore the second $\delta\Sigma$ term. So, in short, the IR-resummation is intended to account for the nonlinear smearing of the BAO signal and is performed in the same way we de-wiggle the power spectrum in Eq.~(\ref{eq:Pdw}). One should note that IR-resummation is not required in LPT.

The state-of-the-art of PT, termed EFTofLSS \citep{D'Amico20}, is to use SPT with EFT and IR-resummation. The publicly available code packages {\sc class-pt} \citep{Chudaykin20} and {\sc pybird} \citep{D'Amico21} allow us to employ the same EFTofLSS model with varying approaches as far as implementation. We choose to work with {\sc pybird}. The EFTofLSS model, which can include a shotnoise term $P_g^{noise}$ but is not here since we remove this in the measurement with {\sc nbodykit}, is as follows: 
\begin{equation}
P(\mathbf{k})|\Theta)
_{EFT}=P^{tree}_g(\mathbf{k})+P^{1-loop}_g(\mathbf{k})+P^{ctr}_g(\mathbf{k}),
\label{eq:Peft}
\end{equation}
where the tree-level term (contains no loops in the Feynman diagram and is what we have considered to be the linear power spectrum) is
\begin{equation}
P^{tree}_g(\mathbf{k}|b_1,f) = \Zcur_1(\mu|b_1,f)P_{lin}(k),
\label{eq:Ptree}
\end{equation}
the 1-loop term is,
\begin{flalign*}
&P^{1-loop}_g(\mathbf{k}|b_1,b_2,b_3,b_4,f) =&\\
&2\int\frac{d^3q}{(2\pi)^3}\Zcur_2(\mathbf{q},\mathbf{k}-\mathbf{q},\mu|b_1,b_2,b_4,f)^2P_{lin}(|\mathbf{k}-\mathbf{q}|)P_{lin}(q)&\\
&+6\Zcur_1(\mu|b_1,f)P_{lin}(k)*&\\
&\int\frac{d^3q}{(2\pi)^3}\Zcur_3(\mathbf{q},-\mathbf{q},\mathbf{k},\mu|b_1,b_2,b_3,b_4,f)P_{lin}(q),&
\end{flalign*}
where $\Zcur_1$, $\Zcur_2$, and $\Zcur_3$ are the redshift-space galaxy density kernels as detailed in Appendix A of \cite{D'Amico21}, and the leading-order quantum corrections to classical field theory is
\begin{flalign*}
&P^{ctr}_g(\mathbf{k}|b_1,f,c_{ct},c_{r,1},c_{r,2},k_m,k_{nl})=2\Zcur_1(\mu|b_1,f)P_{lin}(k)*&\\
&\left(c_{ct}\frac{k^2}{k_{nl}^2}+c_{r,1}\mu^2\frac{k^2}{k_m^2}+c_{r,2}\mu^4\frac{k^2}{k_m^2}\right),&
\label{eq:Pct}
\end{flalign*}
with the scaling parameters $k_m^{-1}$ and $k_{nl}^{-1}$ being the comoving wavelength enclosing the mass of a galaxy and the wavelength indicating non-linear scale, respectively. The coefficient $c_{ct}$ is the matter density counterterm parameter related to the scaling parameter $k_{nl}^{-1}$ and the coefficients $c_{r,1}$, and $c_{r,2}$ are redshift counterterm parameters present with EFTofLSS in redshift-space that are related to the velocity field sampled in the RSD and the scaling parameter $k_m^{-1}$ (see Eq.~(3.9) in \citealt{Perko16}). The parameter convention we use with {\sc pybird} combines the counterterms and scaling parameters as the three free variables $c_1=c_{ct}/k_{nl}^2$, $c_2=c_{r,1}/k_m^2$, and $c_3=c_{r,2}/k_m^2$.
\par

Using {\sc pybird} then entails giving the container a linear power spectrum (produced by either {\sc camb} or {\sc class}) with the corresponding values of $k$, the growth rate $f_g$, and the $7$ EFT parameters $(b_1,b_2,b_3,b_4,c_1,c_2,c_3)$ to then output the nonlinear power spectrum redshift-space monopole $P_0$ and quadrupole $P_2$ for $k_{max}\leq 0.3$, for each redshift slice. 

One is also able to include the AP-effects by first supplying {\sc pybird} with the fiduciary values for the angular diameter distance $D^{fid}_A$ and the Hubble parameter $H^{fid}$ and then giving $D_A$ and $H$ as free variables, where the input of these parameters encoded in {\sc pybird} is in the form of $\tilde{D}_A=D_A(Z)H(z=0)$ and $\tilde{H}=H(z)/H(z=0)$ resulting in the distortion parameters
\begin{equation}
q_\parallel=\frac{H^{fid}(z)/H^{fid}(z=0)}{H(z)/H(z=0)}, \hskip 0.5cm q_\perp=\frac{D_A(z)H(z=0)}{D_A^{fid}(z)H^{fid}(z=0)},
\label{eq:q}
\end{equation}
which have the same dependencies as the geometric distortion parameters $\alpha_\parallel$ and $\alpha_\perp$ since the scale of the sound horizon is 
\begin{equation}
r_s(z_d)=\frac{1}{H(z=0)}\int_{z_d} ^\infty\frac{c_s(z)}{H(z)/H(z=0)}\mathrm{d}z,
\label{eq:rs}
\end{equation}
where $H(z)/H(z=0)=\sqrt{\Omega_{m,0}(1+z)^3+\Omega_{\Lambda,0}}$ and considering that we do not varying $\Omega_{m,0}$ and $\Omega_{\Lambda,0}=1-\Omega_{m,0}$ such that the variable part of $r_s(z_d)$ in $\alpha_{\parallel,\perp}$ is only $H(z=0)$. We can then give {\sc pybird} the inputs for angular diameter distance as $\tilde{D}_A(z)=\alpha_\perp D_A^{fid}(z)H^{fid}(z=0)$ and the input for the Hubble parameter as $\tilde{H}(z)=H^{fid}(z)/H^{fid}(z=0)/\alpha_\parallel$. Likewise, for the growth parameter we can understand that since we give {\sc pybird} the linear power spectrum without leaving $P_0$ as a free variable, the $\alpha_g$ parameter in Eq.~(\ref{eq:growthparam}) reduces to $f_g(z)/f_{g,fid}$ such that we can give the input for the growth rate factor as $\tilde{f}_g=\alpha_g f_{g,fid}$.
\par

For the {\sc pybird} implementation of the EFT model we then have 10 parameters $\Theta=({b_1,b_2,b_3,b_4,c_1,c_2,c_3,\alpha_\parallel,\alpha_\perp,\alpha_g})$ that we find constraints for in an MCMC the same way we do for the phenomenological models with the following prior bounds: $b_1=(0,4)$, $b_{2,3,4}=(-10,10)$, $c_{1,2,3}=(-30,30)$, and $\alpha_{\parallel,\perp,g}=(0.25,1.75)$. We are also able to account for the geometry of the survey with the window function by passing to {\sc pybird} our values for the configuration-space window masks $W_{0,2,4}$ as calculated in Eq.~(\ref{eq:windMLT}). Note that the IR-resummation is also incorporated into the algorithm.

With this, we are able to produce constraints using EFTofLSS as we have for the phenomenological models explored in this work to compare the applicability of these simple models relative to the state-of-the-art PT, in the hope that one can employ a simple model as a valid stand in for a more complete model to be swapped in later, which could be desirable if you were attempting to build tools for bispectrum analysis, for instance, but do not care so much about how the nonlinear power spectrum is produced, just that it be a good prediction of what one is likely to encounter once applied to actual data. Of course, it may be that some particular physical differences arise in the bispectrum which are degenerate in observations of the power spectrum, but this could depend on if one models only the bispectrum monopole or its higher order multipoles. Future work will explore the performance of these phenomenological models and EFTofLSS when used as nonlinear models for bispectrum cosmological analysis (McCarthy, \textit{in prep}).

\begin{table*}
    \caption{Results of comparison statistics (Bayesian Information Criteria (BIC), Figure-of-Bias (FoB), and Figure-of-Merit (FoM)) for $P(k)$ nonlinear modelling combinations that include the {\sc HaloModel} nonlinear correction $\Fcur_{HM}$ for $k_{max}=0.25, 0.3$ and $0.35$ $h$/Mpc. FoM values are given in units of $10^3$. Lowest values of BIC and FoB and highest values of FoM are in bold with second lowest BIC/FoB and second highest FoM in italics.}
    \centering
    \begin{tabular}{lcccccc}
\hline    
\hline 
$\Fcur_{HM}$ Model & BIC & BIC & BIC & FoB(FoM) & FoB(FoM) & FoB(FoM) \\
$1.2<z<1.4$&$k_{max}=0.25$ & $k_{max}=0.3$ & $k_{max}=0.35$ & $k_{max}=0.25$ & $k_{max}=0.3$ & $k_{max}=0.35$ $h$/Mpc \\
\hline 
$P_{lin}*\Fcur_{HM}*\Mcur_A$ & 45.446 & \textbf{49.880} & 61.977 &     2.644(\textbf{77}) & 3.560(\textbf{116}) & 3.754(\textbf{141})\\
$P_{lin}*\Fcur_{HM}*\Mcur_B$ & 45.607 & \textit{50.599} & \textit{59.032} &     1.118(49) & \textit{1.463}(57) & \textit{0.967}(52)\\
\hline 
$P_{dw}(k|k_*)*\Fcur_{HM}*\Mcur_A$  & \textit{44.345} & 50.612 & 63.763 &    2.455(\textit{55}) & 3.387(\textit{95}) & 3.579(70)\\
$P_{dw}(k|k_*)*\Fcur_{HM}*\Mcur_B$  & \textbf{43.697} & 51.342 & \textbf{58.881} &    \textit{0.891}(47) & 1.520(77) & 1.041(\textit{79})\\
\hline
$P_{dw}(k|\Sigma_\perp,\Sigma_\parallel)*\Fcur_{HM}*\Mcur_A$  & 47.890 & 54.585 & 67.329 &      2.167(10) & 2.956(22) & 2.947(17)\\
$P_{dw}(k|\Sigma_\perp,\Sigma_\parallel)*\Fcur_{HM}*\Mcur_B$  & 47.509 & 55.305 & 62.355 &    \textbf{0.694}(19) & \textbf{1.102}(31) & \textbf{0.782}(23)\\
\hline
\hline 
    \end{tabular}
    \label{table:ModCompHM}
\end{table*}

\begin{table*}
    \caption{Results of comparison statistics (Bayesian Information Criteria (BIC), Figure-of-Bias (FoB), and Figure-of-Merit (FoM)) for $P(k)$ nonlinear modelling combinations that include the {\sc galacticus} semi-analytic model nonlinear correction $\Fcur_{SAM}$ for $k_{max}=0.25, 0.3$ and $0.35$ $h$/Mpc. FoM values are given in units of $10^3$. Lowest values of BIC and FoB and highest values of FoM are in bold with second lowest BIC/FoB and second highest FoM in italics.}
    \centering
    \begin{tabular}{lcccccc}
\hline    
\hline 
$\Fcur_{SAM}$ Model & BIC & BIC & BIC & FoB(FoM) & FoB(FoM) & FoB(FoM) \\

$1.2<z<1.4$&$k_{max}=0.25$ & $k_{max}=0.3$ & $k_{max}=0.35$ & $k_{max}=0.25$ & $k_{max}=0.3$ & $k_{max}=0.35$  \\
\hline 

$P_{lin}*\Fcur_{SAM}*\Mcur_A$ & 51.820 & 55.743 & 69.125 &     0.951(\textbf{36}) & 1.105(\textit{65}) & 2.642(\textbf{86})\\

$P_{lin}*\Fcur_{SAM}*\Mcur_B$ & 54.181 & 59.232 & 72.431 &    \textbf{0.345}(15) & 0.457(9) & 1.363(16)\\
\hline 
$P_{lin}*\Fcur_{SAM}*\Mcur_A$ w$Q_3$ & 55.377 & 59.102 & 69.253 &   1.060(\textbf{36}) & 0.917(\textbf{67}) & 2.263(38)\\

$P_{lin}*\Fcur_{SAM}*\Mcur_B$ w$Q_3$ & 57.713 & 62.250 & 71.999 &    \textit{0.397}(16) & \textit{0.383}(14) & \textit{0.712}(5)\\

\hline 
$P_{dw}(k|k_*)*\Fcur_{SAM}*\Mcur_A$ & \textbf{46.564} & \textbf{52.826} & \textbf{65.073} &    1.348(21) & 1.272(49) & 3.243(\textit{61})\\

$P_{dw}(k|k_*)*\Fcur_{SAM}*\Mcur_B$ & 50.172 & 56.948 & 69.081 &    0.596(21) & 0.482(43) & 1.563(42)\\
\hline 
$P_{dw}(k|k_*)*\Fcur_{SAM}*\Mcur_A$ w$Q_3$ & 50.084 & \textit{55.990} & \textit{65.831} &    1.392(\textit{23}) & 1.097(48) & 2.511(47)\\

$P_{dw}(k|k_*)*\Fcur_{SAM}*\Mcur_B$ w$Q_3$ & 53.892 & 60.487 & 69.965 &    0.538(\textit{23}) & \textbf{0.309}(33) & 1.136(46)\\

\hline 
$P_{dw}(k|\Sigma_\perp,\Sigma_\parallel)*\Fcur_{SAM}*\Mcur_A$  & \textit{49.898} & 56.603 & 67.897 &     1.408(11) & 0.983(21) & 2.183(34)\\

$P_{dw}(k|\Sigma_\perp,\Sigma_\parallel)*\Fcur_{SAM}*\Mcur_B$  & 53.564 & 60.736 & 72.280 &     0.742(8) & 0.534(8) & 1.115(15)\\

\hline 
$P_{dw}(k|\Sigma_\perp,\Sigma_\parallel)*\Fcur_{SAM}*\Mcur_A$ w$Q_3$  & 53.492 & 59.814 & 68.839 &     1.447(13) & 0.879(21) & 1.660(28)\\

$P_{dw}(k|\Sigma_\perp,\Sigma_\parallel)*\Fcur_{SAM}*\Mcur_B$ w$Q_3$ & 57.037 & 64.041 & 73.092 &     0.678(6) & 0.443(6) & \textbf{0.679}(5)\\

\hline 
\hline 

    \end{tabular}
    \label{table:ModCompSAM}
\end{table*}

\begin{table*}
    \caption{The results of the BIC statistic for the best choice from nonlinear $P(k)$ modelling combinations with $\Fcur_{HM}$ and with $\Fcur_{SAM}$, from Table~\ref{table:ModCompHM} and Table~\ref{table:ModCompSAM}, respectively, compared to the best-fit BIC values obtained with the Effective Field Theory model described in $\S$\ref{sec:eft} for $k_{max}={0.25,0.3,0.35}$ $h$/Mpc. Note that the EFT model we employ using {\sc pybird} does not allow for computations with $k_{max}>0.3$. Lowest values in bold. }
    \centering
    \begin{tabular}{lcccc}
\hline    
\hline 
BIC $k_{max}=0.25$ [$h$/Mpc] & $1.0<z<1.2$ & $1.2<z<1.4$ & $1.4<z<1.6$ &  $1.6<z<2.0$  \\
\hline 

$P_{dw}(k|k_*)*\Fcur_{HM}*\Mcur_{B}$ & \textbf{56.964} & \textbf{43.697} & \textbf{45.611} & 90.902 \\ 

$P_{dw}(k|k_*)*\Fcur_{SAM}*\Mcur_{B}$ w$Q_3$ & 67.891 & 53.892 & 55.648 & 86.034 \\

$P_{EFT}(k|\psi)$                  & 68.770  &  52.305  &  53.690  &  \textbf{83.505}\\

\hline 
\hline 
BIC $k_{max}=0.3$ [$h$/Mpc] & $1.0<z<1.2$ & $1.2<z<1.4$ & $1.4<z<1.6$ &  $1.6<z<2.0$ \\
\hline 
$P_{dw}(k|k_*)*\Fcur_{HM}*\Mcur_{B}$ & \textbf{67.043} & \textbf{51.342} & \textbf{51.245} & 102.849  \\

$P_{dw}(k|k_*)*\Fcur_{SAM}*\Mcur_{B}$ w$Q_3$ & 74.953 & 60.487 & 61.318 & 94.968 \\ 

$P_{EFT}(k|\psi)$                      & 96.486  &  66.004  &  82.054  &  \textbf{91.269} \\

\hline
\hline 
BIC $k_{max}=0.35$ [$h$/Mpc] & $1.0<z<1.2$ & $1.2<z<1.4$ & $1.4<z<1.6$ &  $1.6<z<2.0$ \\
\hline 
$P_{dw}(k|k_*)*\Fcur_{HM}*\Mcur_{B}$ & \textbf{76.695} & \textbf{58.881} & \textbf{60.140} & 121.685  \\

$P_{dw}(k|k_*)*\Fcur_{SAM}*\Mcur_{B}$ w$Q_3$ & 81.220 & 69.965 & 69.076 & \textbf{104.085} \\ 

$P_{EFT}(k|\psi)$                          & N/A  &  N/A  &  N/A &  N/A \\

\hline
\hline
    \end{tabular}
    \label{table:BIC}
\end{table*}

\begin{table*}
    \caption{The results of the FoB(FoM) statistic for the best choice from nonlinear $P(k)$ modelling combinations with $\Fcur_{HM}$ and with $\Fcur_{SAM}$, from Table~\ref{table:ModCompHM} and Table~\ref{table:ModCompSAM}, respectively, compared to the best-fit BIC values obtained with the Effective Field Theory model described in $\S$\ref{sec:eft} for $k_{max}={0.25,0.3,0.35}$ $h$/Mpc. Note that the EFT model we employ using {\sc pybird} does not allow for computations with $k_{max}>0.3$. Lowest values of FoB and highest values of FoM in bold.
    }
    \centering
    \begin{tabular}{lccccc}
\hline     
\hline 
FoB(FoM) $k_{max}=0.25$ [$h$/Mpc] & $1.0<z<1.2$ & $1.2<z<1.4$ & $1.4<z<1.6$ &  $1.6<z<2.0$  \\
\hline 

$P_{dw}(k|k_*)*\Fcur_{HM}*\Mcur_{B}$ & 2.125(\textbf{11}) & 0.891(\textbf{47}) & 2.518(\textbf{52}) & 1.900(46) \\

$P_{dw}(k|k_*)*\Fcur_{SAM}*\Mcur_{B}$ w$Q_3$ & \textbf{0.824}(5) & \textbf{0.538}(23) & \textbf{1.084}(36) & 1.675(\textbf{47}) \\ 

$P_{EFT}(k|\psi)$                            & 1.346(10)  &  1.550(25)  &  1.883(26)  &  \textbf{1.419}(27) \\

\hline 
\hline 
FoB(FoM) $k_{max}=0.3$ [$h$/Mpc] & $1.0<z<1.2$ & $1.2<z<1.4$ & $1.4<z<1.6$ &  $1.6<z<2.0$ \\
\hline 

$P_{dw}(k|k_*)*\Fcur_{HM}*\Mcur_{B}$ & 1.200(10) & 1.520(\textbf{77}) & 2.615(\textbf{71}) & 2.502(\textbf{83}) \\ 

$P_{dw}(k|k_*)*\Fcur_{SAM}*\Mcur_{B}$ w$Q_3$  & \textbf{0.804}(4) & \textbf{0.309}(33) & \textbf{1.188}(46) & \textbf{1.370}(48) \\

$P_{EFT}(k|\psi)$                            & 1.933(\textbf{15})  &  2.862(38)  &  2.566(32)  &  2.009(35) \\

\hline
\hline 
FoB(FoM) $k_{max}=0.35$ [$h$/Mpc] & $1.0<z<1.2$ & $1.2<z<1.4$ & $1.4<z<1.6$ &  $1.6<z<2.0$ \\
\hline 
$P_{dw}(k|k_*)*\Fcur_{HM}*\Mcur_{B}$ & \textbf{0.595}(\textbf{9}) & \textbf{1.041}(\textbf{79}) & 2.589(\textbf{63}) & 2.516(\textbf{104})  \\

$P_{dw}(k|k_*)*\Fcur_{SAM}*\Mcur_{B}$ w$Q_3$ & 1.709(7) & 1.136(46) & \textbf{1.796}(43) & \textbf{1.313}(35) \\ 

$P_{EFT}(k|\psi)$                            & N/A  &  N/A  &  N/A &  N/A \\

\hline
\hline
    \end{tabular}
    \label{table:FoB}
\end{table*}

\begin{figure*}
\includegraphics[width=\textwidth]{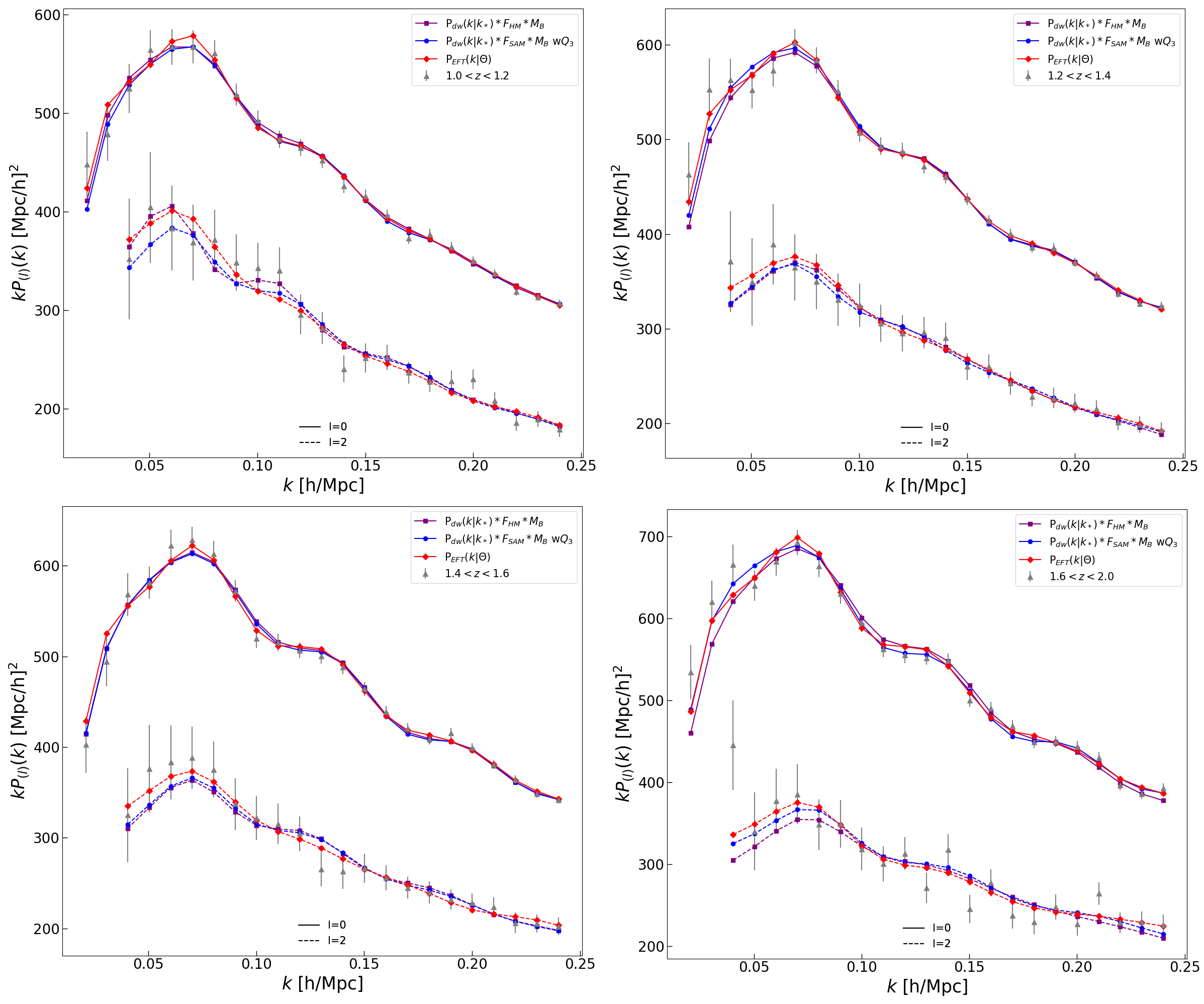}
\caption{The best-fit clustering results from the MCMC chains are shown for the best models from Table's~\ref{table:ModCompHM} and \ref{table:ModCompSAM}, $P_{dw}(k|k_*)*\Fcur_{HM}*\Mcur_B$ in squares and $P_{dw}(k|k_*)*\Fcur_{SAM}*\Mcur_B~wQ_3$ in circles, respectively, for $k_{max}=0.25$ and each redshift slice: $1.0<z<1.2$, $1.2<z<1.4$, $1.4<z<1.6$, $1.6<z<2.0$. We also include a comparison with the EFTofLSS model, $P_{EFT}(k|\Theta)$, in diamonds. Data are shown as triangle points with 1$\sigma$ errors.}
\label{fig:plotBEST}
\end{figure*}

\begin{figure*}
\includegraphics[width=\textwidth]{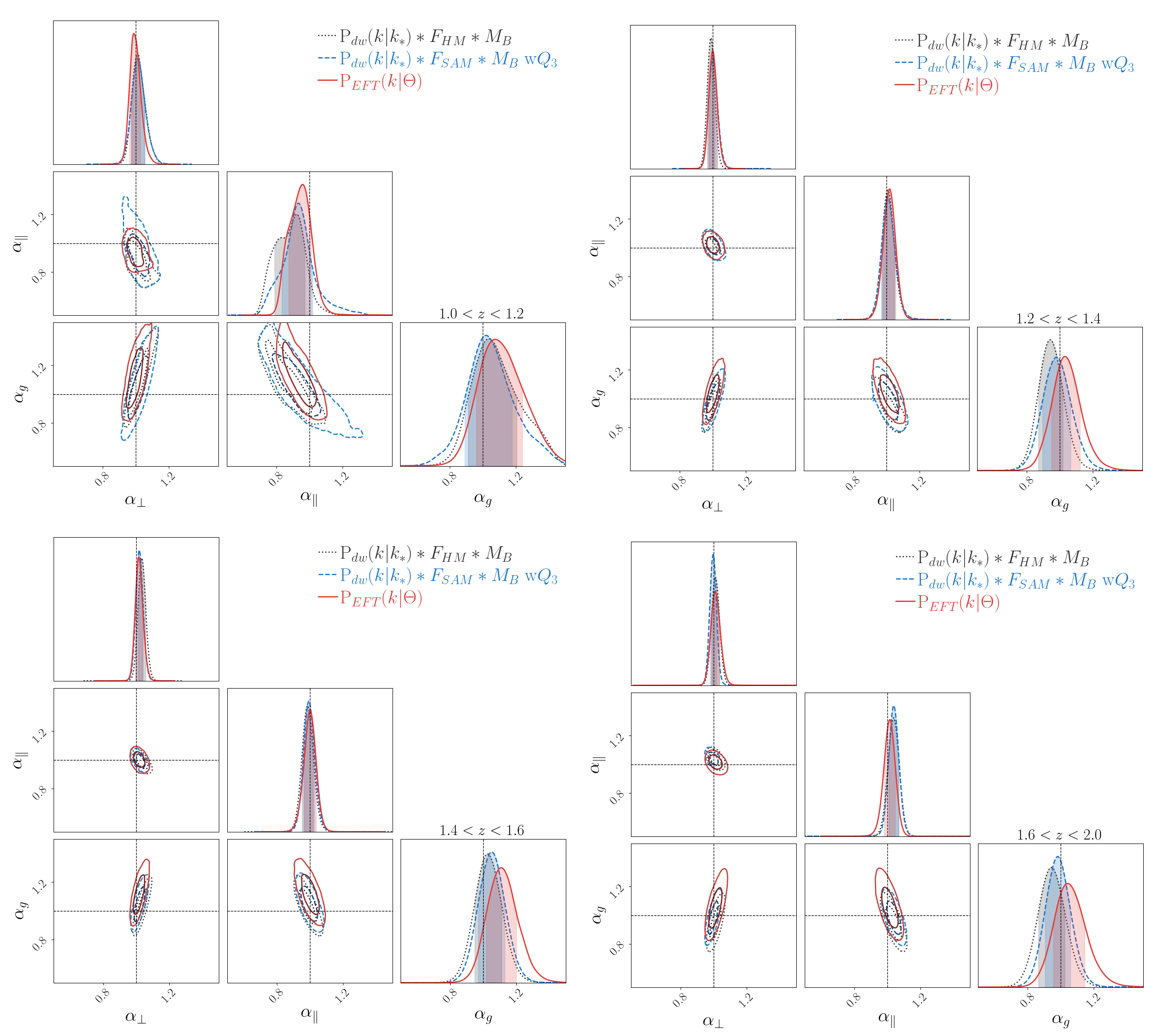}
\caption{Constraints on $\alpha_\perp$, $\alpha_\parallel$, and $\alpha_g$ parameters are shown for the best models from Table's~\ref{table:ModCompHM} and \ref{table:ModCompSAM}, $P_{dw}(k|k_*)*\Fcur_{HM}*\Mcur_B$ in dotted lines and $P_{dw}(k|k_*)*\Fcur_{SAM}*\Mcur_B~wQ_3$ in dashed lines, respectively, for $k_{max}=0.25$ and each redshift slice: $1.0<z<1.2$, $1.2<z<1.4$, $1.4<z<1.6$, $1.6<z<2.0$. We also include a comparison with the EFTofLSS model, $P_{EFT}(k|\Theta)$, in solid lines. Recovery of fiducial cosmology is indicated by the dashed cross hairs in the 2D contours and vertical dashed lines in the margin. }
\label{fig:alphas}
\end{figure*}

\section{Results and Discussion}\label{sec:results}

\subsection{Comparison of different models}
For the comparison between the different combinations of models, it seemed most natural to split them according to the nonlinear corrections $\Fcur_{HM}$ and $\Fcur_{SAM}$ such that the models that include $\Fcur_{HM}$ are presented in Table~\ref{table:ModCompHM} and those that include $\Fcur_{SAM}$ are presented in Table~\ref{table:ModCompSAM}. Previous work \citep{Zhai21} explored modeling the 
nonlinear power spectrum in redshift-space with $P_{dw}(k|k*)$ and $\Fcur_{SAM}$ with either $\Mcur_A$ or $\Mcur_B$, finding that $\Mcur_B$ was a better RSD modeling choice, but did not intend for this to be an exploration of different modeling combinations. Rather, \citealt{Zhai21} served as an introduction to the Roman HLSS {\sc galacticus} galaxy mock, an updated version of which we use in this work to study the clustering of the expected Roman H$\alpha$ galaxies. This work is an extension in that we look to test if that particular combination with $\Fcur_{SAM}$ is preferable to other possible choices and if there are any models with $\Fcur_{HM}$ that are applicable to this redshift range and galaxy target selection.\par

We chose to evaluate all the models for the redshift range $z=1.2-1.4$ because the first redshift slice ($z=1.0-1.2$) was found to be too noisy with jackknife covariances, allowing the models to find local minimums and, often times, having parameters that did not converge in the MCMC. While this was informative to elucidate the limitations and inherent flexibility of the different choices, it is not ideal since with a more careful selection of prior ranges or a better estimation of the covariance we would make different considerations, while the $z=1.2-1.4$ observation allowed all models to work within the same region of the parameter space providing us with a better footing to compare the model combinations. The goal is to identify the best combination with $\Fcur_{HM}$ and the best combination with $\Fcur_{SAM}$ and then compare those over all redshift slices to the EFT model.
\par

In Table~\ref{table:ModCompHM}, we give the BIC, FoB, and FoM (in parenthesis) results for the nonlinear correction $\Fcur_{HM}$ with $P_{lin}$, $P_{dw}(k|k_*)$, or $P_{dw}(k|\Sigma_\perp,\Sigma_\parallel)$ with either $\Mcur_A$ or $\Mcur_B$ for $k_{max}=0.25,0.3,0.35$. The first and second lowest BIC and FoB values are in bold and italics, respectively, while the first and second highest values of FoM are likewise in bold and italics. As mentioned in $\S$\ref{sec:cons}, we want to find a combination of all three quality statistics that suggest a particular model is the simplest and most physically relevant choice, which means we will not necessary choose the model with the most bold values, but rather consider that the BIC should not be too high, indicating the addition of more parameters without a subsequently better clustering fit, while a combination of low FoB with high FoM indicates that the model is not only accurate but also precise.
\par

What is evident from the clustering BIC results is that $P_{dw}(k|k_*)$ is a better choice then $P_{lin}$ and that freeing up the smearing dimensional parameters in $P_{dw}(k|\Sigma_\perp,\Sigma_\parallel)$ did not result in a substantially better clustering fit producing higher BIC values from the additional parameter. When comparing FoB values, we see that the choices with $\Mcur_B$ are always lower, with $P_{dw}(k|\Sigma_\perp,\Sigma_\parallel)$ having the lowest values. However, when we consider the FoM we notice that it is $P_{dw}(k|\Sigma_\perp,\Sigma_\parallel)$ that has the lowest values, which means it is these model combinations that have the largest constraint contours. The highest values of FoM are produced by the $P_{lin}*\Fcur_{HM}*\Mcur_A$ model, a result of having the fewest number of model parameters, while at the same time having lowest FoB values, which means that it is a more precise model but less accurate. The best choice is then $P_{dw}(k|k_*)*\Fcur_{HM}*\Mcur_B$, where we see that the FoB is comparable to $P_{lin}*\Fcur_{HM}*\Mcur_B$ while the FoM is generally higher. Choosing $P_{dw}(k|k_*)*\Fcur_{HM}*\Mcur_B$ rather than $P_{lin}*\Fcur_{HM}*\Mcur_B$ is supported also by the BIC results.
\par

In Table~\ref{table:ModCompSAM}, we present the quality statistics for the model choices with $\Fcur_{SAM}$ in the same way as we did in Table~\ref{table:ModCompHM}, with the additional choice of including $Q_3$ in Eq.~\ref{eq:Fsam}, a modification to the functional form of $\Fcur_{SAM}$ made by \citealt{Sanchez08} to produce better fits at higher $k$'s than $\Fcur_{SAM}$ as presented in \citealt{Cole05} without $Q_3$. When we compare the BIC values between $P_{lin}$, $P_{dw}(k|k_*)$, and $P_{dw}(k|\Sigma_\perp,\Sigma_\parallel)$ we see, as with did for $\Fcur_{HM}$, that $P_{dw}(k|k_*)$ generally has a better BIC. At $k_{max}=0.35$, the addition of $Q_3$ is closer to the models without $Q_3$ than for $k_{max}=0.25$ or $0.3$ indicating that $Q_3$ does perform better at higher $k_{max}$ but does not immediately appear necessary.
\par

Comparing the FoB, we again see that it is the $\Mcur_B$ RSD model that always produces lower values, with $P_{lin}*\Fcur_{SAM}*\Mcur_B$ having generally the lowest values, second lowest across all $k_{max}$'s. When considering FoM, it is the choices with $P_{dw}(k|\Sigma_\perp,\Sigma_\parallel)$ that perform the worst while the combinations with $P_{lin}$ are generally the best. If we look for a compromise, as we did before, i.e., lower FoB with higher FoM, it is $P_{dw}(k|k_*)*\Fcur_{SAM}*\Mcur_B~wQ_3$ that performs the best. We choose this over $P_{dw}(k|k_*)*\Fcur_{SAM}*\Mcur_B$, since all the FoB values are lower with comparable FoM's. Additionally, including $Q_3$ does appear to improve the model with increased $k_{max}$.
\par

The best modeling combination for $\Fcur_{HM}$ is then $P_{dw}(k|k_*)*\Fcur_{HM}*\Mcur_B$ while the best combination for $\Fcur_{SAM}$ is $P_{dw}(k|k_*)*\Fcur_{SAM}*\Mcur_B~wQ_3$. We extend the analysis for these two models over all redshift slices $z=[(1.0,1.2),(1.2,1.4),(1.4,1.6),(1.6,2.0)]$ and for $k_{max}=0.25,0.3,0.35$ $h$/Mpc with the BIC results presented in Table~\ref{table:BIC} and the FoB(FoM) results in Table~\ref{table:FoB}. We also include the quality statistics calculated from fits to the redshift slices and for $k_{max}=0.25,0.3$ $h$/Mpc for the EFTofLSS model detailed in $\S$\ref{sec:eft}. The EFT model as implemented in {\sc pybird} is limited to $k_{max}\leq 0.3$ $h$/Mpc.
\par

Considering the BIC results for these $3$ models in Table~\ref{table:BIC}, we see that for the first three redshift slices, it is $P_{dw}(k|k_*)*\Fcur_{HM}*\Mcur_B$ that performs the best, likely due to having the fewest number of parameters ($8$ verses $10$ in the $P_{dw}(k|k_*)*\Fcur_{SAM}*\Mcur_B~wQ_3$  and $P_{EFT}(k|\Theta)$ models) since we see in Fig.~\ref{fig:plotBEST} that the best fit curves for $k_{max}=0.25\,h$/Mpc are all very similar. For the first redshift slice, $P_{EFT}(k|\Theta)$ performs just as well as $P_{dw}(k|k_*)*\Fcur_{SAM}*\Mcur_B~wQ_3$ if not better, suggesting that if one was to use a model with $10$ parameters, the EFTofLSS might be a better choice than employing $\Fcur_{SAM}$. In fact, for the last redshift slice we see that the EFT model performs the best. This is likely because we have an extended redshift bin for this slice, $\Delta z=0.4$ vs $\Delta z=0.2$ for the lower redshifts, and the $\Fcur_{HM}$ nonlinear correction is determined for a particular redshift slice, here we use $z=1.8$, while $\Fcur_{SAM}$ is free to fit the result of the integration of the redshift range. If we reduce the bin size, we might find similar results to the lower redshift slices, but since the EFT is a physically motivated model, it is able to tease out the correct clustering, mostly from a better recovery of $\alpha_\parallel$ as seen in the bottom right constraint triangle plot of Fig.~\ref{fig:alphas}. For $k_{max}=0.3\,h$/Mpc, the EFT model does not perform as well as it did for $k_{max}=0.25\,h$/Mpc, leaving $\Fcur_{SAM}$ as the better modelling choice.
\par

Considering the FoB(FoM) results in Table~\ref{table:FoB}, it is clear that $P_{dw}(k|k_*)*\Fcur_{SAM}*\Mcur_B~wQ_3$ has the most lowest values of FoB, while the $P_{dw}(k|k_*)*\Fcur_{HM}*\Mcur_B$ model has the best FoM results. The EFT model has FoB's that are comparable to $P_{dw}(k|k_*)*\Fcur_{HM}*\Mcur_B$ but with lower FoM values. If we evaluate the $\alpha_{\perp,\parallel,g}$ contour plots for $k_{max}=0.25\,h$/Mpc in Fig.~\ref{fig:alphas}, we notice that the improvement in FoB for $P_{dw}(k|k_*)*\Fcur_{SAM}*\Mcur_B~wQ_3$ over $P_{dw}(k|k_*)*\Fcur_{HM}*\Mcur_B$ is not simply due to an increase in the area of the contours suggested by the decrease in the FoM since the peaks of the marginalized distributions for $P_{dw}(k|k_*)*\Fcur_{SAM}*\Mcur_B~wQ_3$ actually shift closer to the fiducial values compared to the $\Fcur_{HM}$ combination. For $P_{EFT}(k|\Theta)$, it is the constraints on $\alpha_g$ which display the biggest discrepancy relative to the other two models, generally producing higher values. What is comforting to see here is that all three of these models are able to recover the input cosmology and share similar correlations between $\alpha_{\perp,\parallel,g}$.
\par

\subsection{Comparison with earlier works}

One of the best phenomenological models that we have found, $P_{dw}(k|k_*)*\Fcur_{SAM}*\Mcur_B$, was applied to the BOSS DR10 data by 
\cite{Wang2017}. That work showed that this model is able to recover the input model of the mock catalogs, and lead to accurate and precise measurement of $H(z)r_s(z_d)$, $D_A(z)/r_s(z_d)$, and $f_g(z)\sigma_8(z)$. 

One of the key goals of this paper is to evaluate the uncertainty of BAO and RSD signals from Roman. The simulation and modeling is similar to \citealt{Zhai21}, but with more thorough investigation of the galaxy power spectrum template, thus we are able to recover the input cosmological model more accurately and consistently. When the same model is used, we find consistency of the $\alpha$ scale parameters as expected. However, we should note one difference from the estimate of the covariance matrix. \citet{Zhai21} adopt an EZmock approach and produce thousands of approximate mocks. In this work, we use the data set itself to estimate the covariance matrix with a jackknife subsampling method. 

The constraints found here are less tight than results from the Fisher matrix forecast of \citet{Wang22}, indicating that we can further tighten the constraints by improving the modeling approach. The lack of significant improvement in constraints by going to smaller scales (from $k_{max}=0.25\,h$/Mpc to $k_{max}=0.3\,h$/Mpc to $k_{max}=0.35\,h$/Mpc ) supports this. 
One possibility is to use a suite of tailored galaxy mocks, e.g., from BAM (\citealt{BAM_2019}), to compute the covariance matrix, as jackknife approach is known to overestimate the measurement uncertainties. Although a detailed comparison of different methods is beyond the scope of this paper, we should be aware of the possible impact from the modeling given the high statistical precision expected from Roman HLSS. Examples for the test of covariance matrix along this direction can be found in \citet{Mohammd_2022} and \citet{Percival_2022}.

\section{Conclusion}

In this work, we have explored possible nonlinear corrections to the linear predication of cold dark matter theory, in order to accurately model the galaxy clustering signal in Fourier-space for the Roman High Latitude Spectroscopic Survey \citep{Wang22}. Roman is expected to observe $\sim 10M$ H$\alpha$ emission line galaxies between redshifts $1.0-2.0$, with an H$\alpha$ flux $>10^{-16}$ [ergs/s/cm$^2$] at a signal-to-noise ratio of $6.5\sigma$. Redshifts from [OIII] will also be obtained by Roman at higher redshifts, but we focus on the H$_\alpha$ survey only in this work. \par

We use a lightcone galaxy mock created using the semi-analytical model (SAM) {\sc galacticus} to paint galaxies onto a cosmological N-body simulation. The method requires a tuning of the dust model to 
match observational data. The mock we use adopts the dust model calibrated to the HiZELS H$\alpha$ luminosity function. We make a cut in the H$\alpha$ flux at $>10^{-16}$ [ergs/s/cm$^2$] in redshift slices $z=[(1.0,1.2),(1.2,1.4),(1.4,1.6),(1.6,2.0)]$. We then measure the power spectrum monopole and quadrupole multipole moments following FKP methodology. We utilize the code package {\sc camb} to predict the linear power spectrum from cosmological parameters for the redshift of interest, shift the model into redshift-space with the inclusion of redshift-space distortions, i.e., Kaiser squashing and Finger-of-God effect, and apply a window function to account for the geometry of our simulated survey. \par

We evaluate two methods to account for the nonlinear evolution of the baryons that smear out the BAO signal, a single parameter method P$_{dw}(k|k_*)$ that fixes the ratio between the smearing scale in the perpendicular and parallel dimension and P$_{dw}(k|\Sigma_\perp,\Sigma_\parallel)$ that keeps the smearing scale free in each dimension, and two methods to account for nonlinear structure growth, one that emulates the behavior of the halos seen in N-body simulations with the halo model as its analytical form $\Fcur_{HM}$, and another that emulates the behavior of the galaxies seen in a SAM created with N-body simulations $\Fcur_{SAM}$. We explore combinations of the linear (P$_{lin}$) or de-wiggled linear (P$_{dw}(k|k_*)$ or P$_{dw}(k|\Sigma_\perp,\Sigma_\parallel)$) power spectrum with a growth prefactor ($\Fcur_{HM}$ and $\Fcur_{SAM}$ with or without the addition of $Q_3$). To shift these models into redshift-space, we employ two different techniques to account for the RSD signal: $\Mcur_A$ that assumes the canonical Kaiser squashing and $\Mcur_B$ that includes a window function on $\beta$ in the Kaiser squashing term to isolate the squashing that results from the coherent infall. In all, we explore $6$ different modeling combinations with $\Fcur_{HM}$ and $12$ different combinations with $\Fcur_{SAM}$ for the redshift-space galaxy clustering signal in Fourier-space.

To determine if a particular modeling combination is applicable for the observed clustering signal, we perform a recovery test, keeping the input cosmology fixed and producing constraints on the AP-effect parameters $\alpha_\perp$ and $\alpha_\parallel$ and a similar parameter for the linear growth parameter $\alpha_g$ through application of an MCMC technique. If we have a correct model then $\alpha_{i}=1$ ($i=\perp,\parallel, g$), since the observed cosmology is the same as the fiducial cosmology. We use three quality statistics, the Bayesian Information Criterion (BIC), the Figure-of-Bias (FoB), and the Figure-of-Merit (FoM) to determine which models are the simplest while also being accurate and precise. 

From a careful consideration of these the quality statistics presented for combinations with $\Fcur_{HM}$ in Table~\ref{table:ModCompHM} and for combinations with $\Fcur_{SAM}$ in Table~\ref{table:ModCompSAM} for the redshift slice $z=1.2-1.4$ and $k_{max}=0.25,0.3,0.35\,h$/Mpc, we find that $P_{dw}(k|k_*)*\Fcur_{HM}*\Mcur_B$ and $P_{dw}(k|k_*)*\Fcur_{SAM}*\Mcur_B~wQ_3$ are the best combinations for the respective nonlinear corrections. We then evaluate these two best models over all redshift slices for $k_{max}=0.25,0.3,0.3\,h$/Mpc, giving the BIC results in Table~\ref{table:BIC} and the FoB(FoM) results in Table~\ref{table:FoB}. 

Considering these quality statistics and the best-fit clustering results in Fig.~\ref{fig:plotBEST} and the $\alpha_{\perp,\parallel,g}$ constraints of Fig.~\ref{fig:alphas}, it would seem that $P_{dw}(k|k_*)*\Fcur_{SAM}*\Mcur_B~wQ_3$ is the best at recovering the input cosmology, something we would have expected given that we are modeling the nonlinear power spectrum of a SAM constructed mock with a SAM constructed nonlinear correction, while $P_{dw}(k|k_*)*\Fcur_{HM}*\Mcur_B$ is comparable, suggesting that the nonlinear effects of the galaxies is almost negligible at these redshifts and that a linear galaxy bias is sufficient with a nonlinear model constructed from simulating the behavior of halos. Therefore, if one desires a simple model for quick theoretical exploration of Roman clustering data, $P_{dw}(k|k_*)*\Fcur_{HM}*\Mcur_B$ should suffice as long as one is careful about the parameter priors, covariance calculation, and redshift range explored.
\par

As a sanity check, to verify that these phenomenological models are reasonably physical and applicable to these observations, we include in the comparison of $P_{dw}(k|k_*)*\Fcur_{HM}*\Mcur_B$ and $P_{dw}(k|k_*)*\Fcur_{SAM}*\Mcur_B~wQ_3$ in Table's~\ref{table:BIC} and \ref{table:FoB} the state-of-the-art perturbation theory model EFTofLSS as implemented by {\sc pybird} \citep{D'Amico21}, $P_{EFT}(k|\Theta)$, for $k_{max}=0.25$ and $0.3\,h$/Mpc. 
We find that for the results with $k_{max}=0.25$ [h/Mpc], shown in Fig.~\ref{fig:plotBEST} and \ref{fig:alphas}, the two phenomenological models recover the input cosmology just as well, if not better, than the EFT model, the EFT model generally producing too high values for $\alpha_g$. For $k_{max}=0.3\,h$/Mpc, the EFT model is less able to recover the input cosmology.

This may be an expected trend given that \citealt{D'Amico20} found they could safely perform the analysis of the DR12 BOSS data with negligible theoretical errors only up to $k_{max}=0.2\,h$/Mpc, while \citealt{D'Amico21} found that BOSS pre-reconstructed and post-reconstructed data can be analyzed up to, respectively, $k_{max}=0.23\,h$/Mpc and $k_{max}=0.3\,h$/Mpc. In this work we are modelling the pre-reconstructed signal, meaning that using the EFT model at $k_{max}=0.25\,h$/Mpc may suffer from non-negligible theoretical errors, but we should also considering that the Roman observations are at a higher redshift then the SDSS/BOSS observations where the universe is more linear. We do not model $k_{max}=0.35\,h$/Mpc with $P_{EFT}(k|\Theta)$ since the {\sc pybird} code can only model clustering measurements for $k_{max}\leq 0.3\,h$/Mpc.

Additionally, it has been found that due to non-Gaussianities in the posterior distribution of the counterterm parameters, the marginalized cosmological parameters may be biased by the prior selection \citep{Carrilho23}, which has not been treated with care in our implementation of {\sc pybird}. However, beta testing of a new branch of {\sc pybird} is being tested that is pre-packaged to explore this possibility (Pierre Zhang, \textit{priv. com.}). That being said, this is likely a $<1\sigma$ shift that depends on the total volume of the survey such that if $16x$ the volume of BOSS is observed, i.e., the approximate volumes of DESI and Euclid, the selection of priors is expected to be less informative \citep{Simon22}.
\par

Our current analysis is a purely cosmological study with only a few parameters modeling galaxy formation physics. At large scales, this may be sufficient and the net impact can be described by a single galaxy bias parameter. However, as we go to smaller scales, the impact due to galaxy formation is more complicated and becomes non-negligible (\citealt{McCarthy19}). The {\sc galacticus} SAM mock that we have used provides a useful framework for such exploration given its large volume and galaxy property parameters. 

Our work is an extended analysis using the Roman SAM galaxy mock for the large scale structure analysis based on two point statistics. It is anticipated that constraining power will be significantly enhanced by adding higher order statistics such as galaxy bispectrum. This is our ongoing work, and will be presented elsewhere (McCarthy et. al., in prep).

The High Latitude Wide Area Spectroscopic Survey (HLWASS) that Roman will execute will be determined in an open community process. The HLSS studied here has served as the reference baseline in Roman mission development. We expect that the tools that we have developed in this work, and continue to develop, will be useful in probing dark energy and testing gravity using Roman HLWASS data in an accurate and robust manner.

\section*{Acknowledgments} KSM thanks Dida Markovic and Dan Stern for useful discussions related to the intent of this project, Pierre Zhang and Guido D'Amico for direction regarding the implementation of {\sc pybird}, and the anonymous referee for important revision suggestions that improved the quality of this work. KSM is supported by the NASA Postdoctoral Program. This work was carried out, in part, by IPAC at the California Institute of Technology, and was sponsored by NASA.

\rm{Software used:} Python,
Matplotlib \citep{matplotlib},
NumPy \citep{numpy},
SciPy \citep{scipy},
emcee \citep{ForemanMackey13},
camb \citep{Lewis00},
nbodykit \citep{Hand2018},
pybird \citep{D'Amico21}

\section*{Data Availability}
No new data were generated or analysed in support of this research.

\bibliographystyle{mnras}
\bibliography{paperV}

\bsp    
\label{lastpage}
\end{document}